\documentclass
[onecolumn]{revtex4-1}
\usepackage{subfigure}
\usepackage{amssymb,amsthm,amscd, amsbsy, array}
 \usepackage{amsmath,bm,color}
\usepackage{graphicx} 
\newcommand{\mathsym}[1]{{}}

\usepackage{epsfig}
\usepackage{epsf}    
\usepackage{rotating}
\usepackage{color}

\newcommand{\lf}{\left (}
\newcommand{\lfq}{\left [}

\newcommand{\rg}{\right )}
\newcommand{\rgq}{\right ]}

\def\smallover#1/#2{\hbox{$\textstyle{#1\over#2}$}}
\def\beq{\begin{equation}}
\def\eeq{\end{equation}}
\def\beq{\begin{equation}}
\def\eeq{\end{equation}}
\def\bea{\begin{eqnarray}}
\def\eea{\end{eqnarray}}

\def\lf{\left(}
\def\rg{\right)}

\def\*{{\star}}

\def\xe{{{\'e }}}

\def\aa{{{\`a }}}

\usepackage[authormarkup=none, markup=default, deletedmarkup=xout]{changes}

\definechangesauthor[color=purple]{3}
\definechangesauthor[color=blue]{vito}
\definechangesauthor[color=red]{1}

\begin{document}

 \title{Non-uniform localised distortions in generalised elasticity for liquid crystals}
\author{G. De Matteis \dag \ddag\S \footnote{e-mail: giovanni.dematteis@istruzione.it},  $\quad$  L. Martina \dag \ddag \footnote{e-mail: martina@le.infn.it}, $\quad$ C.  Naya \ddag\footnote{e-mail: carlos.naya@le.infn.it}, $\quad$ V. Turco \dag \ddag\S \footnote{ e-mail: vito.turco@le.infn.it} 
\\ \dag Dipartimento di Matematica e Fisica, Universit\aa del Salento\\  \ddag  INFN, Sezione di  Lecce, Via per Arnesano, C. P. 193 I-73100 Lecce, Italy\\ \S GNFM-INDAM, Citt\`a Universitaria - P.le Aldo Moro 5, C. P. 00185 Roma, Italy \\}

\begin{abstract}
We analyse a recent generalised free-energy for liquid crystals posited by Virga and falling in the class of
quartic functionals in the spatial gradients of the nematic director.
We review some known interesting solutions, {\emph{i. e.}} uniform heliconical structures,  and we find new liquid crystal configurations,
which closely resemble some novel, experimentally detected,  structures called Skyrmion tubes.
These new configurations are characterised by a localised pattern given by the variation of the conical angle. We study the equilibrium differential equations
and find numerical solutions and analytical approximations.
\end{abstract}
  
\maketitle
 
\section{Introduction}

Until recently, uniform nematics, smectics, cholesterics and blue phases have been known to  cover the vast phenomenology observed for liquid crystals \cite{DeGennes,Stewart,RevModPhys46,RevModPhys61}. All of them show interesting new properties and phase transitions when frustrated by geometric confinement and/or external fields \cite{DeGennes,Stewart,RevModPhys46,SelingerDef,Oswald1,Oswald2}. In particular, in recent years it has been demonstrated that chiral nematics can be host for a pletora of new topological and non topological solitonic structures, \emph{i. e.} skyrmions, helicoids, merons and hopfions \cite{TMP18,PRE100,Selinger,Zumer,Smalyukh,SmalyukhPRE}.
Furthermore, skyrmion clusters with mutually orthogonal orientations of the constituent isolated skyrmions have been observed and studied in frustrated chiral liquid crystals \cite{PRB100, PRB98}. 

On the other hand, a new class of nematics has been recently found in bent-core and dimeric systems with clearly recognizable "banana"-like bent-shaped molecules \cite{Chen2013,Borshch2013}.
Besides the conventional uniform nematic $N$ phase, these materials can spontaneously form a new one, now recognized as the {\emph{twist-bend}} nematic phase $N_{TB}$ \cite{Eremin}.
Especially surprising is the fact that the observed new phases exhibit helical (chiral) orientational ordering despite being formed from achiral molecules.
For comparison, more than a century known conventional uniaxial nematic liquid crystals ($N$) are formed from rod-like or disk-like molecules. The chiral cholesteric phases
are locally equivalent to nematics but possess simple (orthogonal) helical structures with pitches in a few micron range: the nematic director $\bm{n}$ twists in space,
drawing a right--angle helicoid and remaining perpendicular to the helix axis. The cholesteric structure appears as a result of relatively weak molecular
chirality (that is why a relatively large pitch), and the swirl direction of the spiral (left or right) is determined by the sign of the molecular chirality.
On the other hand, the action of external fields may induce a deformation of the cholesteric helix into an oblique helicoid with an acute tilt angle\cite{Xiang2014,Xiang2015,Xiang2016,Salili2016,Larentovich2020},
which is usually theoretically analysed on the basis of the classical Frank-Oseen theory.
Unlike this situation, in the $N_{TB}$ nematics the achiral molecules tend to spontaneously, \emph{i. e.} with no external fields,
arrange in heliconical structures with twist and bend deformations: the molecular axes are tilted from the helical axis by an angle $\theta_{TB}$ and the director follows an oblique helicoid.
The $N_{TB}$ phase is stabilized below the uniaxial nematic phase $N$ through both first-order or second-order temperature-driven transitions \cite{greco-SM}, as a result of the spontaneous chiral symmetry breaking. Indeed, a mechanism of this kind is not new to liquid crystal systems (see for instance \cite{Zeldovich1974} or \cite{dzy1}). These structures are similar to the smectic $SmC^*$ phases but, at variance with them, the heliconical textures do not possess any layer periodicity.
Moreover the helical pitch, experimentally found to be on the order of $10\; \text{nm}$, is way smaller than the standard cholesteric one. Direct observation of the periodic heliconical structure was first achieved by the authors of \cite{Chen2013,Borshch2013}
with an estimation of the periodicity around $8\; \text{nm}$.
The new twist-bend nematic represents a structural link between the uniaxial nematic phase $N$, with no tilt, and an unperturbed chiral nematic, {\emph{i. e.}} helicoids with right--angle tilt. 
Although the essential features of the heliconical phase and the $N$-$N_{TB}$ transitions have been outlined by several experimental studies \cite{Salili,Atkinson,Balachandran, Ribeiro,Henderson, Ivsic, Adlem,Sebastian14,Sebastian16}, the elastic properties of the $N_{TB}$ phase still remain unknown.  Nevertheless, several attempts have been made so far to posit a coherent elastic continuum theory \cite{Longa, Barbero1, Barbero2, Barbero3, Virga2, Virga4}.

In fact, by scanning the literature, one can find a number of theoretical works devoted to the twist--bend nematics \cite{Dozov2001, ferrarini, dozov2, selinger2013, Virga2}.
Meyer was the first to hypothesize such heliconical structures in the seventies,  proposing that  they were originated from a spontaneous appearance of the bend flexoelectric polarization \cite{Meyer76}.
Later on, majority of the theoretical works, starting from the influential \cite{Dozov2001}, discuss the question how modulated orientational structures can be formed in achiral systems.
One can easily understand that the description of the twist--bend nematics in terms of an orientational elastic energy requires a pathological (not positively defined) Frank elastic energy.
An analysis in the framework of such Frank energy can explain some experimental observations made for the $\text{N}_{TB}$ liquid crystals,
\emph{e. g.}, anomalously large flexoelectric coefficients \cite{selinger2013} or non-monotonous temperature dependence of the orientational elastic moduli \cite{ferrarini}.
Moreover, in \cite{dzy1} the negative twist elasticity yielding the spontaneous chiral symmetry breaking, has been suggested based on the
Van der Waals contribution into the Frank elastic moduli.
 
In \cite{Dozov2001} Dozov proposed a first elastic theory by using second-order spatial derivatives of the nematic director field,
higher than the first order usually employed in the classical Frank's theory.
Dozov's theory departs from Frank's also for the sign of the bend elastic constant $K_{33}$ which turns negative and, therefore, higher order invariants are needed in the elastic free-energy density in order
to stabilize the heliconical state.
Moreover in \cite{Dozov2001}, Dozov also provides a qualitative description of two $1-$dimensional periodic structures by suitably selecting high order invariants.
For most liquid crystals problems and applications, it is sufficient to consider just the first derivatives of the director. However, in cases such as the twist-bend phase in bent-core liquid crystals or chromonic liquid crystals, where one of the elastic constants may turn negative, higher order terms may have to be added to the Frank-Oseen free energy density. A particular higher-order term has been widely discussed in literature, also concerning its possible role in the stabilization of the twist-bend phase  \cite{Selinger2018}. The Frank-Oseen energy density functional is sometimes written with an additional surface energy contribution besides the saddle-splay term $K_{24}$. It is  a second order surface term called splay-bend, i.e. $K_{13} \, {\rm div}(\boldsymbol n \, {\rm div} \boldsymbol n)$ which was first introduced phenomenologically by Oseen \cite{Oseen1925}, neglected by Frank \cite{Frank1958} and reintroduced by Nehring and Saupe \cite{Nehring1971}. On the other hand, this higher-order term may favour configurations with arbitrary large second derivatives, possibly leading to instabilities.
Later on, several studies addressed the theoretical analysis of $N_{TB}$ phase \cite{selinger2013,Greco,Longa,Virga2}. In particular, in \cite{Greco} a $N$-$N_{TB}$ phase transition was described using
a generalized Maier-Saupe molecular field theory. In \cite{Longa}, a generalized Landau - De Gennes theory was used to investigate 1-dimensional modulated nematic structures generated by  non-chiral and intrinsically  chiral V-shaped molecules. In \cite{Virga2}, the $N_{TB}$ phase was treated as a mixture of two different ordinary $N$ phases, both presenting heliconical structures with the same pitch but opposite helicities. A quadratic elastic theory, still featuring four Frank's elastic constants, was used for each of the two helical phases. Similar models were proposed in \mbox{\cite{Barbero1,Barbero2, Barbero3}}, where also the effects of an external magnetic/electric bulk field were investigated.  Again, authors in \cite{VirgaTandF,Dozov2016} proposed coarse-grained elastic models, which similarly to the model for $SmA^*$ \cite{DeGennes}, make use of an extra scalar order parameter. 
In \cite{Barbero1} $\text{N}_{TB}$ phase elasticity with {\emph{two director fields}}
has been discussed within the positively defined conventional Frank energy. In the recent paper \cite{shamid_selinger_allender} the authors consider
how flexoelectricity combined with spontaneous polar order (ferroelectricity) could stabilize conical spiral orientational order.
However, under natural Landau theory assumptions the model in \cite{shamid_selinger_allender} yields strongly biaxial and polar features of the
$\text{N}_{TB}$ phase, apparently not supported by experimental observations.
In \cite{kats_lebedev} a Landau phenomenological theory was proposed for the phase transition from the conventional nematic phase to
the heliconical phase. The authors of \cite{kats_lebedev} introduce a double-scale elasticity energy by splitting the director fields into two components:
a long-scale Frank energy for a component
of the director and a short-scale elastic energy for the remaining component.

As mentioned above, in his seminal paper \cite{Dozov2001} Dozov proposed higher spatial derivatives of the director nematic field to stabilize his elastic model and to bind the energy from below when $K_{33}$ turns negative. However, there is another way to develop higher order field  theories, that is looking for invariants expressed through higher powers of first derivatives. The latter approach has certainly been applied with success in several field theories, as for example the Skyrme model and the 3-dimensional Skyrme-Faddeev model \cite{Skyrme,Faddeev,Manton,skyrmefad}. It is in this perspective that, very recently, Virga proposed a new fourth order generalized elastic theory for nematics \cite{Virga4} with six elastic constants: three coming from the standard Frank energy second order terms and other additional three associated with fourth order terms.
There it was shown how, for a certain choice of two model parameters, two families of uniform distortions with opposite chirality, exhausting the heliconical structures of the $N_{TB}$ phase, minimize the proposed higher order elastic free-energy. 

In the present paper, after reviewing in a slightly different approach the main results obtained by Virga,
we find new localised solutions for the generalized elastic free-energy posited in \cite{Virga4}. 
On the experimental side, evidence of similar new localised configurations, namely skyrmions, has been found in \cite{PRB100, PRB98}, where the authors showed that the existence of either a conical or uniform state surrounding isolated skyrmions leads to an attracting/repulsive inter-skyrmion potential, respectively. These soliton-like structures in a conical or helical background also appear in ferromagnets. Indeed, not only skyrmions but also the so-called heliknotons
have been recently investigated both in liquid crystals and ferromagnets \cite{heliknotons1,heliknotons2}.  

The rest of the paper is organized as follows. In Sections \ref{secttwo} and \ref{sectthree} we obtain the main results of Virga's work \cite{Virga4}, by expressing
the generalized elastic free-energy in terms of the quantities $(\bm{n},\nabla\bm{n},\text{div}\bm{n},\text{curl}\bm{n})$
and by analysing Meyer's heliconical configurations of $N_{TB}$ phase.
We show how a uniform heliconical state corresponds to a minimum of the fourth order energy density functional for suitable choices of the six elastic constants characterizing Virga's model. 
In Section \ref{sectfour}, we approach the problem of searching for non-uniform distortions departing from the uniform heliconical state, finding localised solutions
similar to those studied in \cite{Dereck} for the Skyrme-Faddeev model. Finally, we draw our conclusions in Section \ref{sectfive} and we suggest possible extensions of the results here presented. 

\section{Mathematical Framework}
\label{secttwo}

Traditionally, nematic liquid crystals are modelled by a general quadratic form in the spatial gradients $\nabla\bm{n}$ of a unit vector, the nematic director $\bm{n}$.
This quadratic form is usually known as Frank's elastic energy density and is written as follows \cite{book_virga}
\begin{equation}
\label{frank_energy_std}
F_{\text{F}} = \frac{1}{2}K_{11}(\text{div}\bm{n})^2+\frac{1}{2}K_{22}(\bm{n}\cdot\text{curl}\bm{n})^2+\frac{1}{2}K_{33}|\bm{n}\times \text{curl}\bm{n}|^2+
K_{24}\left[\text{tr}(\nabla\bm{n})^2-(\text{div}\bm{n})^2\right],
\end{equation}
where $K_{11}, K_{22}, K_{33}$, and $K_{24}$ are Frank's elastic constants. The term $K_{24}$ is a {\emph{null Lagrangian}},
it can be integrated over the domain $\mathcal{B}$ occupied by the nematic medium, without producing any contribution
to the total free-energy provided that $\bm{n}$ is assigned over the boundary $\partial\mathcal{B}$.
As customary, the general formula is often reduced to the one-constant approximation, which can be obtained
by setting $K_{11}=K_{22}=K_{33}=K$ and $K_{24}=\frac{1}{2}K$, thus leading to
\begin{equation}
F_{\text{F}} = \frac{1}{2}K||\nabla\bm{n}||^2,
\end{equation}
where $||\nabla\bm{n}||^2=\partial_in_k\partial_in_k$.
In \cite{Selinger2018}, a new interpretation of (\ref{frank_energy_std}) has been proposed and analysed in depth.
The starting point of this revisited version of Frank's free-energy density formula is the decomposition of $\nabla\bm{n}$
in a set of specific distortion modes.
More precisely, the gradient of $\bm{n}$ can be decomposed as follows
\begin{equation}
\label{decomposition}
\nabla\bm{n}=-\bm{b}\otimes\bm{n}+\frac{1}{2}T\mathbf{W}(\bm{n})+\frac{1}{2}S\mathbf{P}(\bm{n})+\mathbf{D},
\end{equation}
where the scalar $S=\text{div}\bm{n}$ is the {\emph{splay}}, the pseudo-scalar $T=\bm{n}\cdot\text{curl}\bm{n}$ is the {\emph{twist}}, and the vector
$\bm{b}=\bm{n}\times\text{curl}\bm{n}$ is the {\emph{bend}}.
$\mathbf{W}(\bm{n})$ denotes the skew-symmetric tensor associated with $\bm{n}$, {\emph{i. e.}} $\mathbf{W}_{ij}=\epsilon_{ijk}n_k$
and $\mathbf{P}(\bm{n})=\mathbf{I}-\bm{n}\otimes\bm{n}$ is the projector onto the plane orthogonal to $\bm{n}$.
$\mathbf{D}$ is a symmetric traceless tensor such that $\mathbf{D}\bm{n}=\bm{0}$.
Accordingly, it can be given the form
\begin{equation}
\mathbf{D}=q(\bm{n}_1\otimes\bm{n}_1-\bm{n}_2\otimes\bm{n}_2),
\end{equation}
where $q$ is the positive eigenvalue of $\mathbf{D}$ and $\bm{n}_1$ and $\bm{n}_2$ are the eigenvectors, orthogonal to $\bm{n}$. 
From (\ref{decomposition}) it follows that
\begin{equation}
\label{identity1_0}
\text{tr}\mathbf{D}^2=2q^2=\text{tr(}\nabla\bm{n})^2+\frac{1}{2}T^2-\frac{1}{2}S^2.
\end{equation}
In coordinates, we can rewrite the director gradient $\nabla\bm{n}$ as follows \cite{Selinger2018} 
\begin{eqnarray}
\partial_jn_i = -b_in_j+\frac{1}{2}T\epsilon_{ijk}n_k+\frac{1}{2}S\left(\delta_{ij}-n_in_j\right)+D_{ij},
\end{eqnarray}
and $\mathbf{D}$ can be given the alternative forms
\begin{eqnarray}
D_{ij} = \frac{1}{2}\left[\partial_in_j+\partial_jn_i-n_in_k\partial_kn_j-n_jn_k\partial_kn_i-\delta_{ij}\text{div}\bm{n}+n_in_j\text{div}\bm{n}\right],
\end{eqnarray}
or
\begin{eqnarray}
D_{ij} = \frac{1}{2}\left[\partial_in_j+\partial_jn_i+n_ib_j+n_jb_i-S(\delta_{ij}-n_in_j)\right].
\end{eqnarray}
The quantity $q$ was named by Selinger \cite{Selinger2018} as {\emph{biaxial splay}}.
The quantities $(S,T,\bm{b},\mathbf{D})$ are independent from one another and are called {\emph{measures of distortion}}.
Frank's elastic free-energy density can be written as a quadratic form in the four above quantities as follows
\begin{eqnarray}
\label{frank_free_energy}
F_{\text{F}} = \frac{1}{2}\left(K_{11}-K_{24}\right)S^2 + \frac{1}{2}\left(K_{22}-K_{24}\right)T^2+\frac{1}{2}K_{33}B^2+K_{24}\text{tr}(\mathbf{D}^2),
\end{eqnarray}
where $B^2=\bm{b}\cdot\bm{b}$. As recalled in \cite{Virga4}, the positive definiteness of the quadratic form (\ref{frank_free_energy}) implies that
\begin{equation}
K_{11}-K_{24}>0,\qquad K_{22}-K_{24}>0,\qquad K_{33}>0,\qquad K_{24}>0,
\end{equation}
known as Ericksen's inequalities \cite{ericksen1966}. Accordingly, (\ref{frank_free_energy}) admits as global minimiser the state
\begin{equation}
\label{UniformState}
S = T = B = q = 0,
\end{equation}
which corresponds to any constant field $\boldsymbol{n} \equiv \boldsymbol{n_0}$.

As pointed out by authors in \cite{Selinger2018}, there may be a further surface term in (\ref{frank_energy_std}), which, in contrast with the saddle-splay term, i.e. $K_{24}$ term, contains second derivatives of the director. This term reads as ${\rm div}(\boldsymbol n \, {\rm div} \boldsymbol n)=S^2+(\boldsymbol n \cdot \nabla)S$. In analogy with what is done for the $K_{24}$ contribution, one can in principle expand $\partial_i \partial_j n_k$ in its normal modes and build a generalised energy density functional with also the second order part expressed in terms of this modes. However, as recalled by \cite{Virga4} one may obtain a generalized elastic free energy density by introducing higher powers in the expansion of the gradient of $\boldsymbol n$. As we will show, these higher order terms, together with the negative sign of $K_{33}$, are sufficient to accommodate the twist-bend phase as ground state and no second derivative term, as the $K_{13}$, is actually needed.

The eigenvectors $\bm{n}_1,\bm{n}_2,\bm{n}$ of $\mathbf{D}$ are called the {\emph{distortion frame}}. This can be defined
for any sufficiently regular director field $\bm{n}$ and it changes from point to point, thus defining a movable frame.
The bend vector $\bm{b}$ can be decomposed in the distortion frame as follows
\begin{equation}
\bm{b} = b_1\bm{n}_1+b_2\bm{n}_2.
\end{equation}
The scalars $(S,T,b_1,b_2,q)$ depend on position in space and they are called collectively {\emph{distortion characteristics}} of the nematic director.
In \cite{Virga4} it was introduced the concept of {\emph{uniform distortion}}, \emph{i. e.} a uniform distortion is a configuration
of the director field $\bm{n}$ in which the distortion characteristics are the same everywhere, whilst the distortion frame may change from place to place.
Any constant field $\bm{n}\equiv \bm{n}_0$ is uniform but with no distortion, thus it is clear that a uniform distortion should also have a non-trivial director pattern. 
In \cite{Virga4}, it was shown that there exist only two families of uniformly distorted director fields and they are given by
\begin{equation}
\label{uniformdistortion1}
S=0,\qquad T=2q, \qquad b_1=b_2=b,
\end{equation}
\begin{equation}
\label{uniformdistortion2}
S=0, \qquad T=-2q,\qquad b_1=-b_2=b,
\end{equation}
where, of course, $q$ and $b$ are constant assigned parameters. In order to reconstruct the structure of the director fields $\bm{n}$ corresponding
to (\ref{uniformdistortion1}) and (\ref{uniformdistortion2}), one has to integrate the decomposed spatial gradient (\ref{decomposition}) with
the specific distortion characteristics given by (\ref{uniformdistortion1}) and (\ref{uniformdistortion2}).
In \cite{Virga4}, it was shown that the most general uniform distortion is an {\emph{heliconical}} director field, more precisely a director field of the form
\begin{equation}
\label{heliconics_1}
\bm{n}_h=\sin\theta_0\cos\beta z\bm{e}_x+\sin\theta_0\sin\beta z\bm{e}_y+\cos\theta_0\bm{e}_z,
\end{equation}
where $\bm{e}_x,\bm{e}_y,\bm{e}_z$ are the cartesian unit  basis vectors in $\mathbb{R}^3$, and the {\emph{conical}} angle $\theta_0$ and the pitch $2\pi/|\beta|$ are related to the parameters $(q,b)$ as follows
\begin{equation}
\label{parameters_heliconics}
\frac{2\pi}{|\beta|}=\frac{2\pi q}{b^2+2q^2},\qquad\qquad \cos\theta_0 = \frac{|b|}{\sqrt{b^2+2q^2}}.
\end{equation}
The nematic director $\bm{n}$ rotates around $\bm{e}_z$ making a fixed cone angle $\theta_0$ with the rotation axis $\bm{e}_z$, which is called the helix axis.
The distortion frame $\left\{\bm{n}_1,\bm{n}_2,\bm{n}\right\}$ processes along $\bm{e}_z$ turning completely round over the length of a pitch $2\pi/|\beta|$
and it remains unchanged in all directions orthogonal to $\bm{e}_z$.
The structure (\ref{heliconics_1}) describes therefore the heliconical distortion predicted by Meyer \cite{Meyer76} and it corresponds to the {\emph{twist-bend}}
liquid crystal phase, experimentally detected in 2011 \cite{cestari}. It is worth noticing that formula (\ref{heliconics_1}) also describes  the nematic phase $N$
when $\theta_0=0$ and the chiral nematics when $\theta_0=\frac{\pi}{2}$, implying that the twist-bend phase represents a structural link between these two extreme phases.

Of course, as also observed in \cite{Virga4} the heliconical configurations cannot be minimizers of the standard Frank elastic energy
and it is needed a new elastic theory able to accommodate the heliconical phase as a ground state.
In \cite{Virga4}, it was put forward a new energy functional with quartic powers of the spatial gradient of $\bm{n}$.
The starting point for positing the quartic energy functional is the set of measures of distortion $(S, T, \bm{b},\mathbf{D})$.
In order to form a quartic polynomial in the spatial gradients of $\bm{n}$ we need to collect
the basic invariants under nematic symmetry $\bm{n}\leftrightarrow-\bm{n}$, rotations and inversions, {\emph{i. e.}}
\begin{equation}
\left\{S^2,T^2,B^2,\text{tr}\mathbf{D}^2, S\bm{b}\cdot\mathbf{D}\bm{b},T\bm{b}\cdot\mathbf{D}(\bm{n}\times\bm{b})\right\}.
\end{equation}
From the list above it would be possible to construct a general higher order polynomial. However, we shall follow the approach in \cite{Virga4}
and we will consider the minimalistic quartic free-energy density as follows
\begin{equation}
\label{virga_free_energy}
F_{TB}(S,T,b_1,b_2,q)=\frac{1}{2}k_1S^2+\frac{1}{2}k_2T^2+k_2\text{tr}\mathbf{D}^2+\frac{1}{2}k_3B^2+\frac{1}{4}k_4T^4+k_4(\text{tr}\mathbf{D}^2)^2+\frac{1}{4}k_5B^4+
k_6T\bm{b}\cdot\mathbf{D}(\bm{n}\times\bm{b}).
\end{equation}
This represents the lowest order free-energy density that, for a suitable choice of the elastic constants, admits as global minimizer the heliconical uniform distortion state (\ref{heliconics_1}), characterized by (\ref{uniformdistortion1}) and (\ref{uniformdistortion2}), as opposed to the uniform state (\ref{UniformState}). Notice that other quartic contributions are not included as this is the simplest way to guarantee that the global minimum is attained at (\ref{uniformdistortion1}) and (\ref{uniformdistortion2}).

By taking into account that $\text{tr}\mathbf{D}^2=2q^2$ and
\begin{equation}
\bm{b}\cdot\mathbf{D}(\bm{n}\times\bm{b}) = -2qb_1b_2,
\end{equation}
formula (\ref{virga_free_energy}) can also be written as a function of the characteristics of distortion
\begin{equation}
F_{TB}(S,T,b_1,b_2,q)=\frac{1}{2}k_1S^2+\frac{1}{2}k_2\left[T^2+(2q)^2\right]+\frac{1}{2}k_3B^2+\frac{1}{4}k_4\left[T^4+(2q)^4\right]+\frac{1}{4}k_5B^4-k_6(2q)Tb_1b_2.
\end{equation}
By directly comparing (\ref{virga_free_energy}) with (\ref{frank_free_energy}) we get the following formal identification
\begin{equation}
k_1=K_{11}-K_{24},\quad k_2=K_{22}-K_{24}=K_{24},\quad k_3=K_{33},
\end{equation}
but as shown below $k_3$ can also assume negative values.
The above energy density turns out to be coercive provided that
\begin{equation}
k_4>0,\quad k_5>0,\quad k_6>0,\quad k_6^2<2k_4k_5,
\end{equation}
which is the condition of positive definiteness of the quartic part of (\ref{virga_free_energy}).
Moreover, since the heliconical states (\ref{uniformdistortion1}) and (\ref{uniformdistortion2}) are characterized by $S=0$
we assume $k_1>0$ so that $F_{TB}$ attains its minimum for $S=0$.
Having fixed conditions on the elastic constants $k_1,k_4,k_5,k_6$, it makes sense to classify the minimizers in terms of the remaining $k_2$ and $k_3$ constants.
It was shown in \cite{Virga4} that (\ref{virga_free_energy}) is minimized by the trivial uniform state $\bm{n}=\bm{n}_0$ if and only if $k_3\geq0$. In terms
of the measures of distortions this minimizer is given by (\ref{UniformState}).

When
\begin{equation}
-2\frac{k_5}{k_6}k_2<k_3<0,
\end{equation}
the minimizer is a pure bend state with
\begin{equation}
S=T=q=0,\qquad\qquad B^2=b_1^2+b_2^2=-\frac{k_3}{k_5}.
\end{equation}
Finally, when
\begin{equation}
k_3<-2\frac{k_5}{k_6}k_2<0,
\end{equation}
$F_{TB}$ is minimized by the pure heliconical state
\begin{equation}
\label{heliconics_explicit}
T^2=(2q)^2=-\frac{k_3k_6+2k_5k_2}{2k_4k_5-k_6^2}\geq 0,\qquad\qquad\text{and}\qquad\qquad b_1^2=b_2^2=-\frac{k_2k_6+k_3k_4}{2k_4k_5-k_6^2}\geq 0.
\end{equation}
Summing up, in order to have the heliconical states \eqref{heliconics_explicit}, the following constraints on the elastic constants must hold
\begin{equation}
\label{elastic_constraints}
2 k_4 k_5-k_6^2>0,\qquad k_3 k_6+2k_5 k_2<0,\qquad k_2 k_6+k_3 k_4<0. 
\end{equation}
In the next section we will work out this minimizer by studying the Euler-Lagrange equations associated to $F_{TB}$.
To this aim, it is convenient to write $F_{TB}$ in terms of the spatial gradient components $\nabla\bm{n}$
and of the quantities $\text{div}\bm{n}$, $\text{curl}\bm{n}$. For this we need some identities. It can be proved that
\begin{equation}
2Tqb_1b_2 = (\bm{n}\cdot\text{curl}\bm{n})\text{curl}\bm{n}\cdot\nabla\bm{n}(\bm{n}\times\text{curl}\bm{n})+\frac{1}{2}(\bm{n}\cdot\text{curl}\bm{n})^2|\bm{n}\times\text{curl}\bm{n}|^2.
\end{equation}
Upon using the latter identity along with the definition of $S,T$ and $\mathbf{D}$ and the identity (\ref{identity1_0}), one arrives at
\begin{eqnarray}
\label{virga_free_energy2}
F_{TB}&=&\frac{1}{2}(k_1-k_2)(\text{div}\bm{n})^2+k_2(\bm{n}\cdot\text{curl}\bm{n})^2+k_2\text{tr}(\nabla\bm{n})^2+\frac{1}{2}k_3|\bm{n}\times\text{curl}\bm{n}|^2+
\frac{1}{4}k_4(\bm{n}\cdot\text{curl}\bm{n})^4\nonumber\\
&+&
k_4\left[\text{tr}(\nabla\bm{n})^2
+\frac{1}{2}(\bm{n}\cdot\text{curl}\bm{n})^2-\frac{1}{2}(\text{div}\bm{n})^2\right]^2+\frac{1}{4}k_5|\bm{n}\times\text{curl}\bm{n}|^4
\nonumber\\
&-&k_6\left[(\bm{n}\cdot\text{curl}\bm{n})\text{curl}\bm{n}\cdot(\nabla\bm{n})(\bm{n}\times\text{curl}\bm{n})+
\frac{1}{2}(\bm{n}\cdot\text{curl}\bm{n})^2|\bm{n}\times\text{curl}\bm{n}|^2\right].
\end{eqnarray}
Otherwise, $F_{TB}$ can be rewritten as follows
\begin{eqnarray}
F_{TB} &=& \frac{1}{2}(k_1-k_2)S^2+k_2T^2+k_2\text{tr}(\nabla\bm{n})^2+\frac{1}{2}k_3B^2+\frac{1}{4}k_4T^4+k_4\left[\text{tr}(\nabla\bm{n})^2+\frac{1}{2}T^2-\frac{1}{2}S^2\right]^2
\nonumber \\
&+&\frac{1}{4}k_5B^4-k_6I_{4a}-\frac{1}{2}k_6T^2B^2,
\end{eqnarray}
where
\begin{equation}
I_{4a}=\left[(\bm{n}\cdot\text{curl}\bm{n})\text{curl}\bm{n}\cdot(\nabla\bm{n})(\bm{n}\times\text{curl}\bm{n})\right].
\end{equation}
Correspondingly, the stored free-energy in a region $\mathcal{B}$ is given by the volume integral
\begin{equation}
\label{free_energy_volume}
\mathcal{F}=\int_{\mathcal{B}}{F_{TB}\text{d}\mathcal{B}}.
\end{equation}

\section{Uniform heliconical distortions}
\label{sectthree}

\begin{figure}[ht]
\begin{center}
\includegraphics[width=0.45\textwidth]{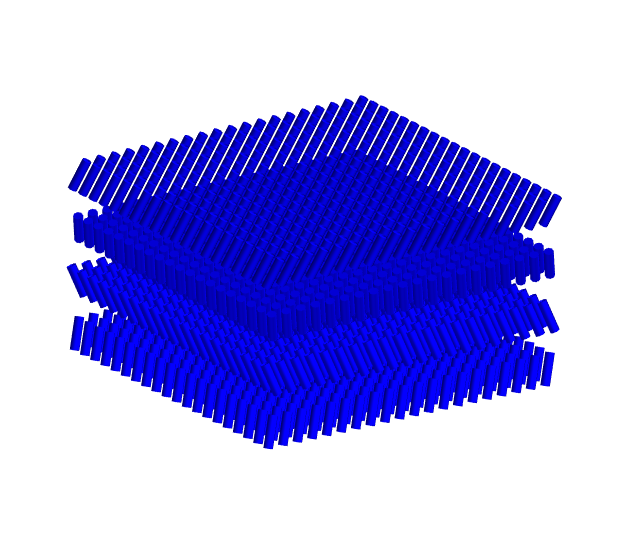}
\includegraphics[width=0.45\textwidth]{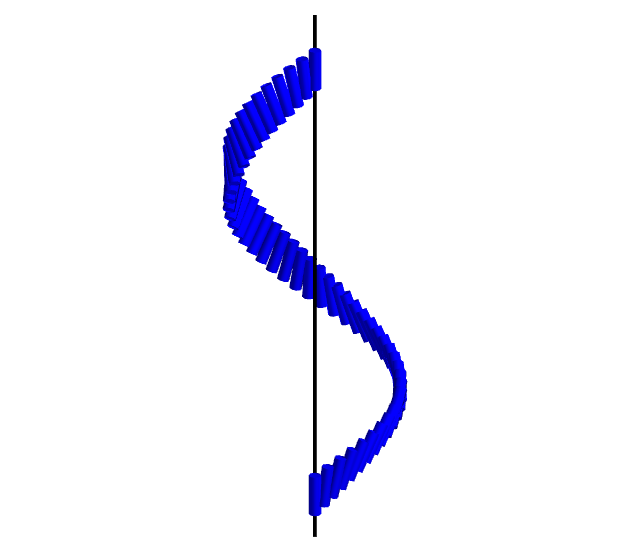}
\end{center}
\caption{Three-dimensional representation of the uniform heliconical distortion. {\it Left:} Different $(x,y)$-plane cross sections showing the change of orientation along the $z$ direction. {\it Right}: Helix line along the $z$ axis for a fixed
distance from it.}
\label{3D-alhpa0}
\end{figure}

In this section we analyse the special class of solutions (\ref{uniformdistortion1})-(\ref{uniformdistortion2}) which, as shown above, provides a global minimum
to the free-energy density functional (\ref{virga_free_energy2}). To this aim, we use the standard parametrization of $\bm{n}$
\begin{equation}
\bm{n} = \sin\theta\cos\phi\bm{e}_x+\sin\theta\sin\phi\bm{e}_y+\cos\theta\bm{e}_z,
\end{equation}
where $\theta$ and $\phi$ are the standard polar angle functions.\\
The general Euler-Lagrange equations, also supplemented with boundary conditions on $\partial\mathcal{B}$, are rather involved.
However, it can be shown that they admit the heliconical configurations \eqref{heliconics_1} as solutions (see \cite{SupMat}). The three-dimensional representation of such configurations is displayed in Fig. \ref{3D-alhpa0},
where a set of $(x,y)$-plane cross sections showing how the configuration changes along $z$ and a specific helix line are depicted.
The corresponding free-energy density reads
\begin{eqnarray}
F_{TB}(\bm{n}_h)=f_{TB}(\theta_0,\beta)&=&\frac{1}{8}\left(-4k_6\cos^2\theta_0\sin^6\theta_0+4k_4\sin^8\theta_0+\frac{1}{8}k_5\sin^42\theta_0\right)\beta^4\nonumber\\
&+&\frac{1}{8}\left(8k_2\sin^4\theta_0+k_3\sin^22\theta_0\right)\beta^2,
\end{eqnarray}
which depends on the pitch-related parameter $\beta$ and the conical angle $\theta_0$, and is minimised by the values
\begin{equation}
\label{beta}
\beta=\pm\frac{(2k_2k_5+k_3k_6)+2(k_3k_4+k_2k_6)}{\sqrt{-(2k_2k_5+k_3k_6)(2k_4k_5-k_6^2)}},
\end{equation}
and
\begin{equation}\label{def_theta_0}
\theta_0=\arcsin{\left(\sqrt{\frac{2k_2k_5+k_3k_6}{(2k_2k_5+k_3k_6)+2(k_3k_4+k_2k_6)}}\right)},
\end{equation}
with the following condition on the elastic constants
\begin{equation}
\label{cond_t_new}
0<\frac{2k_2k_5+k_3k_6}{(2k_2k_5+k_3k_6)+2(k_3k_4+k_2k_6)}<1,
\end{equation}
fully satisfied by the constraints \eqref{elastic_constraints}.
It is worth noticing that both $\theta_0$ and $\beta$ do not depend on the elastic constant $k_1$.
Correspondingly, the free-energy density on the heliconical solutions becomes
\begin{equation}
\label{fenergy_new}
f_{TB}(\theta_0,\beta)=-\frac{1}{2}\frac{k_3^2k_4+2k_2^2k_5+2k_2k_3k_6}{2k_4k_5-k_6^2},
\end{equation}
which, taking into account \eqref{elastic_constraints}, turns out to be negative provided that
\begin{equation}
\label{quadratic_new}
k_3^2k_4+2k_2^2k_5+2k_2k_3k_6>0.
\end{equation}
Actually, the quadratic form (\ref{quadratic_new}) with respect to $k_3$,
\begin{equation}
k_3^2k_4+2k_2^2k_5+2k_2k_3k_6,
\end{equation}
is positive as the discriminant
\begin{equation}
4k_2^2k_6^2-8k_4k_2^2k_5=4k_2^2(k_6^2-2k_4k_5)<0,
\end{equation}
as a consequence of (\ref{elastic_constraints}). Thus, the value of the free-energy density at the heliconical state turns out to be lower than the value at the uniform nematic phase.

Finally, it is possible to verify that the heliconical configurations found here correspond to those predicted in \cite{Virga4} in the form (\ref{heliconics_explicit}).
Actually, by direct computation, using the expressions for $\theta_0$ and $\beta$ above, one can show that
\begin{equation}
S=0,\qquad T^2=\sin^4 \theta_0 \; \beta^2=-\frac{2k_2k_5+k_3k_6}{(2k_4k_5-k_6^2)},\qquad\text{and}\qquad B^2 = b_1^2+b_2^2 = -\frac{2(k_3k_4+k_2k_6)}{2k_4k_5-k_6^2},
\end{equation}
which reproduce formulas (\ref{heliconics_explicit}).

\section{Non-uniform localised states}
\label{sectfour}

\subsection{Ansatz on the solution}

At variance with the previous section, here we consider the case of non-uniform distortions, possibly leading to localised states.
Bearing in mind that the unifom distortions are heliconical states, we slightly depart from this case by considering still heliconical structures,
but with a non-uniform conical angle and an additional precession around the axis of the uniform heliconical state.
More precisely, we consider configurations of the general form
\begin{equation}
\label{general_ansatz}
\bm{n}( r, z,\varphi;\alpha,\beta)=\sin(f(r))\cos(\alpha \varphi+\beta z)\bm{e}_x+\sin(f(r))\sin(\alpha \varphi+\beta z)\bm{e}_y+\cos(f(r))\bm{e}_z,
\end{equation}
where $\alpha$ is an integer describing the number of windings performed by the director around the heliconical axis $\bm{e}_z$ for fixed $z$,
$f(r)$ is the profile function describing the conical angle and
$\beta$ has the same meaning as in the previous section. In order to have localised configurations, we may impose the boundary conditions $f(0)=0$ and $f(r\to\infty)=f_0$, $f_0$ being a suitable conical angle to be determined.
Then, to study these configurations, we need to reduce the general free-energy in order to translate the ansatz into the equilibrium equations.
The reduced free-energy integrated over the unit cell $\left[0,\frac{2\pi}{\beta}\right]\times\left[0,2\pi\right]$ and over $r\in\left[0,\infty\right]$ will take the form
\begin{equation}
\mathcal{F}\left[f;\alpha,\beta\right]=\int_{0}^{\frac{2\pi}{\beta}}{\text{d}z}\int_{0}^{2\pi}{\text{d}\varphi}\int_{0}^{\infty}F_{TB}\left[\bm{n}(r, z,\varphi;\alpha,\beta)\right]r\text{d}r.
\end{equation}
We are interested in the reduced free-energy per unit cell $\left[0,\frac{2\pi}{\beta}\right]\times\left[0,2\pi\right]$
which can be obtained by dividing by the factors $2\pi$ and $\frac{2\pi}{\beta}$
\begin{equation}
\tilde{\mathcal{F}}\left[f;\alpha,\beta\right]=\frac{\beta}{4\pi^2}\mathcal{F}\left[f;\alpha,\beta\right].
\end{equation}
In the following, we will study two relevant cases: $\alpha=0$ and $\alpha=1$.

\subsection{Case $\alpha=0$}

In this first case, we are interested in studying whether localised solutions without  winding around the heliconical axis $\bm{e}_z$ are possible. In fact, this is equivalent to take $\alpha = 0$
in the general ansatz (\ref{general_ansatz}) with a radial dependent profile, $f(r)$, for the conical angle. With these assumptions, the reduced free-energy reads:
\begin{equation}
\label{free_energy_alphazerozero}
\mathcal F_0[f;\beta]=\int_0^{\infty} \left[\Gamma_0(f) + \Gamma_2(f) f'^2 + \Gamma_4(f) f'^4  \right] r dr,
\end{equation}
where the quantities $\Gamma_i$ are functions of the profile $f(r)$, $\beta$ and the elastic constants (see Appendix \ref{app:equation} for details). Hence, the Euler-Lagrange equation is given by
\begin{equation}
\label{alphazeroode}
2 f'' (\Gamma_2 + 6 f'^2 \Gamma_4) + \frac{2}{r} f' \Gamma_2 + f'^2 \partial_f \Gamma_2 + \frac{4}{r} f'^3 \Gamma_4 + 3 f'^4 \partial_f \Gamma_4 - \partial_f \Gamma_0 = 0,
\end{equation}
with $\partial_f \Gamma_i$ the partial derivative of the quantity $\Gamma_i$ with respect to the conical function $f$. 
It is worth noticing that the above equilibrium differential equation is invariant under the following transformation
\begin{equation}
\label{symmetry}
f\rightarrow -f\quad\text{and}\quad r\rightarrow -r.
\end{equation}
First of all, we find the asymptotic state as $r\rightarrow\infty$. To this aim, we take the limit of (\ref{alphazeroode}) as $r\rightarrow\infty$ and we get the asymptotic stationary condition
\begin{equation}
\label{con_infinito_alphazero}
\partial_f \Gamma_0 = 0.
\end{equation}
This last equation corresponds to a stationary condition for the energy functional (\ref{free_energy_alphazerozero}) in the same limit, $r\rightarrow\infty$, where $f^{\prime}$ also vanishes.
Correspondingly, the free-energy per unit cell can be written as
\begin{equation}
\label{free_energy_infinity_alfazero}
\mathcal{\tilde{F}}_0\left[f;\beta\right] = \frac{\beta}{4\pi^2}\int \Gamma_0(f)  r dr+ \text{h.o.t.}\;.
\end{equation}
It is clear that in order to find the corresponding asymptotic state we will need to minimize the leading term of (\ref{free_energy_infinity_alfazero})
with respect to $f$ and $\beta$. Thus, in addition to the condition (\ref{con_infinito_alphazero}) we need to include the stationary condition with respect to $\beta$, {\emph{i. e.}}
letting
\begin{equation}
\label{67_function}
F_{0\infty}(f,\beta,r):=\frac{\beta}{4\pi^2}  \Gamma_0 ( f)r,
\end{equation}
we then have to require that
\begin{equation}
\nabla_{(f,\beta)}F_{0\infty} (f,\beta,r)=(0,0).
\end{equation}
Letting $\tau:=\cos 2f$, (\ref{67_function}) can be written as
\begin{eqnarray}
&&F_{0\infty}(\tau,\beta,r)=\nonumber\\
&&r\frac{1}{64} \beta^2 (\tau-1) \left[\beta^2 (\tau-1) \left(2 k_4 (\tau-1)^2+(\tau+1) (k_5 t+k_5+2 k_6 \tau-2 k_6)\right)+16 k_2 (\tau-1)-8 k_3 (\tau+1)\right].
\end{eqnarray}
The corresponding stationary conditions with respect to $\beta$ and $\tau$ read as
\begin{eqnarray}
\frac{1}{16} \beta^2 \left(\beta^2 (\tau-1) \left(2 k_4 (\tau-1)^2+k_5 \tau (\tau+1)+k_6 (2 \tau+1) (\tau-1)\right)+8 k_2 (\tau-1)-4 k_3 \tau\right)=0,
\end{eqnarray}
\begin{eqnarray}
\frac{1}{32} \beta (\tau-1) \left(2 \beta^2 (\tau-1) \left(2 k_4 (\tau-1)^2+(\tau+1) (k_5 \tau+k_5+2 k_6 \tau-2 k_6)\right)+16 k_2 (\tau-1)-8 k_3 (\tau+1)\right)=0.
\end{eqnarray}
Upon solving them simultaneously we get solutions $\tau_0,\beta$ as follows
\begin{eqnarray}
\tau_0=\frac{-2 k_2 k_5+2 k_2 k_6+2 k_3 k_4 - k_3 k_6}{2 k_2 (k_5+k_6)+k_3 (2 k_4+k_6)},
\end{eqnarray}
\begin{eqnarray}
\label{beta_carlos}
\beta = \pm\frac{(2k_2k_5+k_3k_6)+2(k_3k_4+k_2k_6)}{\sqrt{-(2k_2k_5+k_3k_6)(2k_4k_5-k_6^2)}}.
\end{eqnarray}
The asymptotic conical angle is then given by
\begin{equation}
\label{104_eq_f0}
f_0 = \frac{1}{2}\text{arccos}\tau_0=\frac{1}{2}\text{arccos}\left({\frac{-2 k_2 k_5+2 k_2 k_6+2 k_3 k_4 - k_3 k_6}{2 k_2 (k_5+k_6)+k_3 (2 k_4+k_6)}}\right),
\end{equation}
and is equal to that found in the uniform heliconical state $\theta_0$ (\ref{def_theta_0}).

To find localised solutions, we need to solve the Euler-Lagrange equation (\ref{alphazeroode}) with some specific boundary conditions. In particular, we ask the profile function to reach the asymptotic value of the conical angle $\theta_0$ at infinity,
{\emph{ i. e.}}, $f(r) \rightarrow \theta_0$ when $r \rightarrow \infty$, while at the origin it takes a different value which we will choose as zero for simplicity. 

Unfortunately, after checking numerically, we did not find any localised local minimum.
Thus, the $\alpha=0$ case just reproduces the uniform heliconical state (\ref{heliconics_1}) with a constant conical angle $\theta_0$ as global minimizer.
Nevertheless, it seems reasonable to think that the absence of a winding around a given axis makes it difficult to stabilise solutions interpolating different values of the conical angle. Indeed, additional energy is not needed in varying the value of $f(r)$ at the origin and taking the conical angle $\theta_0$ everywhere, arriving at the state corresponding to the global minimum. However, if the system needs to go through an unwinding before reaching the uniform heliconical distortion, then stable local minima may be allowed. This is in fact what happens when $\alpha=1$, so we will devote the rest of the paper to its study and description.

\subsection{Case $\alpha=1$}

When $\alpha=1$ the reduced free-energy takes the following form
\begin{eqnarray}
\label{free_energy_deflat}
\mathcal{F}_1\left[f;\beta\right] =
\frac{\pi^2}{64 \beta} \int_0^{\infty} \left[ G_0 (r, f) + G_1 (r, f) f' + G_2 (r, f) f'^2 + G_3 (r, f) f'^3 + G_4 (r, f) f'^4 \right] dr,
\end{eqnarray}
and the associated Euler-Lagrange equation turns into an ordinary differential equation of second order of the form
\begin{equation}
\label{eq:ode1}
2f^{\prime\prime}\left(G_2+3f^{\prime}G_3+6f^{\prime 2}G_4\right)+2f^{\prime}\partial_rG_2+f^{\prime 2}\partial_fG_2+2f^{\prime 3}\left(\partial_fG_3+2\partial_rG_4\right)+3f^{\prime 4}\partial_fG_4-
\partial_fG_0+\partial_rG_1=0.
\end{equation}
The quantities $G_i$, $i = 0, 1, 2, 3, 4$ depend on $r, f, \beta, k_1, k_2, k_3, k_4, k_5, k_6$ and are listed below in Appendix A.
Also in this case, it is worth noticing that the above equilibrium differential equation is invariant under the following transformation
\begin{equation}
\label{symmetry}
f\rightarrow -f\quad\text{and}\quad r\rightarrow -r.
\end{equation}
We will use this symmetry property in the following subsections, starting with the investigation of the asymptotic behaviour of the profile function $f(r)$ around $r=0$ and around $r=\infty$.

\subsubsection{Asymptotics}

In order to study the behaviour around $r=0$ we first fix the leading order power at the origin by assuming
that, close to $r=0$, the profile function $f$ takes the form 
\begin{equation}
\label{ex_78}
f(r)=ar^l+\text{O}(r^{l+1}),
\end{equation}
with $l>0$, as a negative value would imply loss of regularity in $f$ at the origin.
In addition, $l$ has to take an odd value due to the symmetry given by Eq. (\ref{symmetry}).  Thus, by expanding the l.h.s. term of (\ref{eq:ode1}) around $f=0$ which is the value taken by $f$ at $r=0$, it can be shown (see \cite{SupMat}) that $l=1$.

Having fixed this leading power, we can now study some relationships among the derivatives of $f$ at $r=0$ by
inspecting the corresponding expansion at $r=0$.
From the above mentioned symmetry property (\ref{symmetry}) of (\ref{eq:ode1}) and from the results about the leading order at $r=0$ we
conclude that the power expansion of $f$ around $r=0$ takes the form 
\begin{equation}
\label{symmetry_exp}
f(r) = \xi r + \zeta r^3+ \eta r^5 +...\;,
\end{equation}
where
\begin{equation}
\label{xizeta}
\xi=f^{\prime}(0),\qquad\quad \zeta = \frac{1}{3!}f^{\prime\prime\prime}(0),\qquad\quad \eta = \frac{1}{5!}f^{\rm (V)}(0)...\;.
\end{equation}
By replacing $f(r)$ with (\ref{symmetry_exp}) in the Euler-Lagrange equation (\ref{eq:ode1}) we get an expansion in the even powers of $r$ only.
In particular, after a lengthy calculation, we arrive at
\begin{eqnarray}
\label{rela_first_third}
\zeta = \frac{3\beta^2k_3 \xi+2(k_1-3k_2-3\beta^2k_6) \xi^3}{12(k_1+3k_2+9k_4 \xi^2)},
\end{eqnarray}
\begin{eqnarray}
&&\eta=\frac{1}{360 \left(k_1+3 k_2+9 k_4 \xi^2\right)}
\Bigg(30 \beta^4 k_5 \xi^3+120 \beta^2 k_2 \xi^3-80 \beta^2 k_3 \xi^3+30 \beta^2 k_3 \zeta+90 \beta^2 k_5 \xi^5+270 \beta^2 k_6 \xi^5-360 \beta^2 k_6 \xi^2 \zeta\nonumber\\
&-&13 k_1 \xi^5+240 k_1 \xi^2 \zeta
+51 k_2 \xi^5-15 k_3 \xi^5-180 k_3 \xi^2 \zeta+48 k_4 \xi^7+1080 k_4 \xi^4 \zeta\nonumber\\
&-&4320 k_4 \xi \zeta^2-30 k_5 \xi^7+45 k_6 \xi^7+360 k_6 \xi^4 \zeta\Bigg).
\end{eqnarray}
This result has been successfully used as a check of the numerical calculations presented below by taking the value of $\xi$ coming from the simulations.

Next we collect the results about the asymptotic analysis as $r\rightarrow\infty$ of equation (\ref{eq:ode1}).
From the functions $G_i$ reported in Appendix \ref{app:equation}, it is not difficult to recognize that the only
surviving term is
\begin{equation}
\label{con_infinito}
\partial_f G_{0}^{\infty}=0,
\end{equation}
where the function $G_0^{\infty}$ is obtained from the function $G_0$ by dropping
all the terms $1/r$ and $1/r^3$ and keeping only linear terms in $r$, {\emph{i. e.}
\begin{eqnarray}
G_{0}^{\infty} = g_{01}^{\infty} + g_{02}^{\infty}\cos 2 f+ g_{03}^{\infty}\cos 4 f + g_{04}^{\infty}\cos 6 f + g_{05}^{\infty}\cos 8 f,
\end{eqnarray}
where
\begin{eqnarray}
g_{01}^{\infty}& =& \frac{\beta^2}{2}r(192k_2+32k_3+70\beta^2k_4+3\beta^2k_5-10\beta^2k_6),
\end{eqnarray}
\begin{eqnarray}
g_{02}^{\infty}&=&-4\beta^2 r(32k_2+14\beta^2k_4-k_6\beta^2),
\end{eqnarray}
\begin{eqnarray}
g_{03}^{\infty}&=&+2\beta^2 r(16k_2-8k_3+14\beta^2 k_4-\beta^2 k_5+2\beta^2 k_6),
\end{eqnarray}
\begin{eqnarray}
g_{04}^{\infty}&=&-4\beta^4 r (2k_4+k_6),
\end{eqnarray}
\begin{eqnarray}
g_{05}^{\infty}&=&\frac{\beta^4}{2}(2k_4+k_5+2k_6)r,
\end{eqnarray}
entailing that
\begin{equation}
g_{01}^{\infty}+g_{02}^{\infty}+g_{03}^{\infty}+g_{04}^{\infty}+g_{05}^{\infty} = 0.
\end{equation}
Correspondingly, from (\ref{free_energy_deflat}),  the free-energy per unit cell in the same limit reduces to
\begin{eqnarray}
\label{free_energy_infinity}
\tilde{\mathcal{F}_1}\left[f;\beta\right] =
\frac{1}{256} \int \left[ G_0^{\infty} (r, f) \right] dr + \text{h.o.t.}\;.
\end{eqnarray}
Taking into account the expressions for $\Gamma_0$ in Appendix \ref{app:equation} and $G_0^{\infty}$, it should be noticed that the expression for $\tilde{\mathcal{F}_1}\left[f;\beta\right]$ is the same as the one obtained for $\tilde{\mathcal{F}_0}\left[f;\beta\right]$
in (\ref{free_energy_infinity_alfazero}). Hence we obtain identical stationary conditions and the same asymptotic state, as far as the conical angle $f_0$ and $\beta$
are concerned (see eqs. (\ref{beta_carlos}) for $\beta$ and (\ref{104_eq_f0}) for $f_0$). As for $\alpha=0$ the expression for $f_0$ reproduces the same as that for $\theta_0$ in (\ref{def_theta_0}).
Moreover, in order to avoid divergences at infinity of the free-energy density, we also need to subtract (see sections above) the free-energy density
value for the uniform heliconical configuration (general global minimum) from the general free-energy expression. This actually corresponds to the case $\alpha=0$.

Finally, it is worth noting that the profile $f(r)$ approaches the asymptotical conical angle by means of a modified Bessel function of the second kind and order zero, namely,
\begin{equation}
\label{perturb}
f(r) = f_0+\epsilon h(r), \qquad h(r)=c_2K_0(\omega r)\approx c_2\sqrt{\frac{\pi}{2}}\frac{e^{-\omega r}}{\sqrt{\omega r}}+...\;,
\end{equation}
where $c_2$ is an arbitrary constant and $\omega$ is a parameter coming from the Euler-Lagrange equation and depending on the elastic constants only (see \cite{SupMat}).

\subsubsection{Global approximation: Pad\'e approach}

In the previous subsection we have analysed the asymptotic behaviour of the solution of the equation  
\eqref{eq:ode1}
near the boundaries  0 and $+ \infty$.  Now,  one may try to look for an analytic expression, which approximates the true solution in some specific sense.
To this aim, we take inspiration from similar second--order ODEs involving trigonometric non-linearities, like the simple pendulum equation or the challenging $P_{\text{III}}$ Painlev\'e equation (see for instance \cite{website}).
First, one may apply a suitable transformation in terms of inverse trigonometric functions of the dependent variable, leading to a rational expression of the  equation in the new dependent variable and its derivatives.
Then, one may  more easily study  and possibly obtain a suitable approximated solution,  with the method adopted for instance in \cite{MMZ}. Thus, we look for a solution of the form 
 \beq  f\lf r \rg = \pi - \arccos\lfq s\lf r \rg \rgq, \label{trasftrig}\eeq
 where $s\lf r \rg$ is an unknown function subject to the conditions
 \beq \lim_{r \to 0^+}s\lf r \rg =  -1 , \qquad  \lim_{r \to +\infty}s\lf r \rg = - \cos\lf f_0\rg , \label{limiting}\eeq
$f_0$ being defined by the asymptotic value (\ref{104_eq_f0}). Indeed, the adopted transformation (\ref{trasftrig}) maps \eqref{eq:ode1}  into an ODE involving only algebraic rational expressions, {\emph{i. e.}} 
combinations of powers of $s\lf r \rg$ and its derivatives, up to the second order, with non constant coefficients. However,  these coefficients can be  Laurent expanded  in the neighborhood of the boundaries.
Accordingly, one may guess that also the function $s\lf r \rg$ could be expressed  as a ratio of polynomials in $r$, possibly of infinite degree. Moreover, having already noted above that $f\lf r \rg$ must contain only odd powers of $r$,
it follows from the properties of the $\arccos$ function,  that $s\lf r \rg$ must be an even function of $r$.
As a consequence, $s(r)$ and its Taylor expansion must depend on $r^2$ only.  Now, it is well known that Pad\'e approximants are a powerful tool to study the convergence of given Taylor series and they are exact on rational functions \cite{BGrMo}.
To this end, let $ S_N = \sum_{j = 0}^N c_j \; r^j $ be a truncated series  at the order $N$ of our function $s(r)$,
its  Pad\'e approximant $s^{\lfq L / M\rgq}$  of order $\lf L, M \rg , \; L + M = N , $ is given by
\beq   \label{pincopadina} s^{\lfq L / M\rgq} = \frac{\sum_{j = 0}^L a_j \; r^j}{\sum_{j = 0}^M b_j \; r^j},  \qquad b_0 = 1,\eeq
such that \beq S_N - s^{\lfq L / M\rgq} = \text{O}\lf r^{N+1} \rg. \label{approxi} \eeq
In our case we have a boundary value problem with two different series expansions at $r=0$ and $r=\infty$, respectively, which have to be joined simultaneously by the searched $s^{\lfq L / M\rgq}$.
This is a well known problem in  the multipoint Pad\'e approximation, which could be solved in terms of continuous fractions  (see \cite{BGrMo} Vol 2). However, in the present context, one may proceed in
a more straightforward way as follows.
 
First,  by  power expanding around  $r=0$ the function $f\lfq s\lf r \rg \rgq$ and comparing it with formulas (\ref{symmetry_exp}) and (\ref{xizeta}),  one obtains the corresponding expansion for $s(r)$
 \beq  s\lf r \rg =-1+\frac{r^2 \xi^2}{2}+\frac{1}{24} r^4 \left(24 \xi \zeta-\xi^4\right)+\frac{1}{720} r^6 \left(720 \eta \xi+\xi^6-120 \xi^3 \zeta+360
   \zeta^2\right)+\text{O}\left(r^7\right) . \label{h6}\eeq
On the other hand, we need a  similar expansion at infinity, for which we use the simplest asymptotic expansion
\beq \label{naive_expa} s\lf r \rg = - \cos\lf f_0\rg  +\text{O}\lf \frac{1}{r^2}\rg.
\eeq
Assuming $f_0 \neq \frac{\pi}{2}$,  a possible Pad\'e approximant (\ref{pincopadina}) must  have  equal highest powers in both numerator and denominator.  Thus, one may  choose $L = M = 4$,  and then set 
\beq s^{\lfq 4/4\rgq}  = \frac{a_2 r^4+a_1 r^2+a_0}{b_2 r^4+b_1 r^2+1},  \label{appr44}\eeq
where the five constants  $a_i$ and $b_i$  have to be determined by using the information contained in both asymptotic expansions above. 
This can be done by formally expanding $s^{\lfq 4/4\rgq}$ around $r=0$ and $r=\infty$ and by matching the corresponding coefficients with those in (\ref{h6}) and (\ref{naive_expa}).
We stop at the fourth order in (\ref{h6}) and accordingly, the coefficient  $\eta$, being involved at the sixth order,  will not be considered in further calculations.
This procedure leads to finding three relations  providing the coefficients $a_i$, namely
\beq  a_0=-1,\quad a_1=\frac{\xi^2}{2}-b_1, \quad a_2=\frac{1}{24} \left(12 b_1 \xi^2-24 b_2-\xi^4+24
   \xi \zeta\right). \label{aiii}\eeq The constant $b_1$ can  be determined by  resorting to \eqref{naive_expa}
and using  (\ref{aiii}) to obtain
  \beq  b_1 =  \frac{-24 b_2 \cos f_0 +24 b_2+\xi^4-24 \xi \zeta}{12 \xi^2} .\eeq
 Finally, using the above expressions into $s^{\lfq 4/4\rgq} $ , one is led to the approximation
  \beq f\lf r \rg = \pi -\arccos\left(\frac{-12 b_2 r^4 \xi^2 \cos f_0+r^2 \left(24 b_2
   \cos f_0-24 b_2+5 \xi^4+24 \xi \zeta\right)-12 \xi^2}{r^2 \left(-24 b_2
   \cos f_0+24 b_2+\xi^4-24 \xi \zeta\right)+12 b_2 r^4 \xi^2+12 \xi^2}\right) . \label{fapp4}\eeq
 Thus, we are restrained to  choose the four parameters  $\xi, \zeta, b_2, f_0$.  The most obvious choice for $f_0$  is to use its expression given by (\ref{104_eq_f0}) in terms  of the elastic constants.  Secondly,  the quantity $\zeta$  can  be  expressed in terms of $\xi$ and of the elastic constants by (\ref{rela_first_third}).  Thus, one can  consider the family of functions depending only on  the couple $\lf \xi, b_2\rg$, which can be determined by  a  best fit (in the sense of the minimum squares method) with the numerical solution of the differential equation \eqref{eq:ode1}.   
A detailed discussion of these methods and their results is contained in the next section. 

\subsubsection{Numerical analysis}

Due to the non-linearity and complexity of the system under consideration, also the use of numerical methods seems mandatory.
In the following, we want to numerically study localised solutions of the form (\ref{general_ansatz}),
with $\alpha = 1$.
Here, for notational convenience, we will employ a different symbol for the stored free-energy $\mathcal{F}_1$ in (\ref{free_energy_deflat}), {\emph{i. e.}}
\begin{equation}
E_{\alpha=1} = \frac{\pi^2}{64 \beta} \int \left[ G_0 (r, f) + G_1 (r, f) f' + G_2 (r, f) f'^2 + G_3 (r, f) f'^3 + G_4 (r, f) f'^4 \right] dr,
\end{equation}
where the integration over $r$ is being performed on a finite domain (a similar notation, \emph{i. e.} $E_{\alpha=0}$, is used for the stored free-energy $\mathcal{F}_0$ in (\ref{free_energy_alphazerozero})).
Hence, to find configurations minimising this energy we use a gradient flow method in one dimension applied to a lattice of 1000 points with an interspace of $\Delta r = 0.02$. In addition, spatial derivatives are approximated by a finite fourth-order accurate difference. The values of the profile function $f(r)$ at the boundaries of the grid are $f(0) = 0$ and $f(r \rightarrow \infty) = \theta_0$, see equation (\ref{def_theta_0}).

The behaviour at the origin comes from the fact that if we want the field $\boldsymbol n$ to be well-defined, $f(0)$ has to be zero or an integer multiple of $\pi$. Indeed, the condition $f(0) = \pi$ has been also considered but it poses a higher energy with respect to the vanishing profile. This might be expected since it implies a bigger deviation from the global minimum given by the uniform distortion $f = \theta_0$.
\begin{figure}[ht] 
\begin{center}
\includegraphics[width=.7\textwidth]{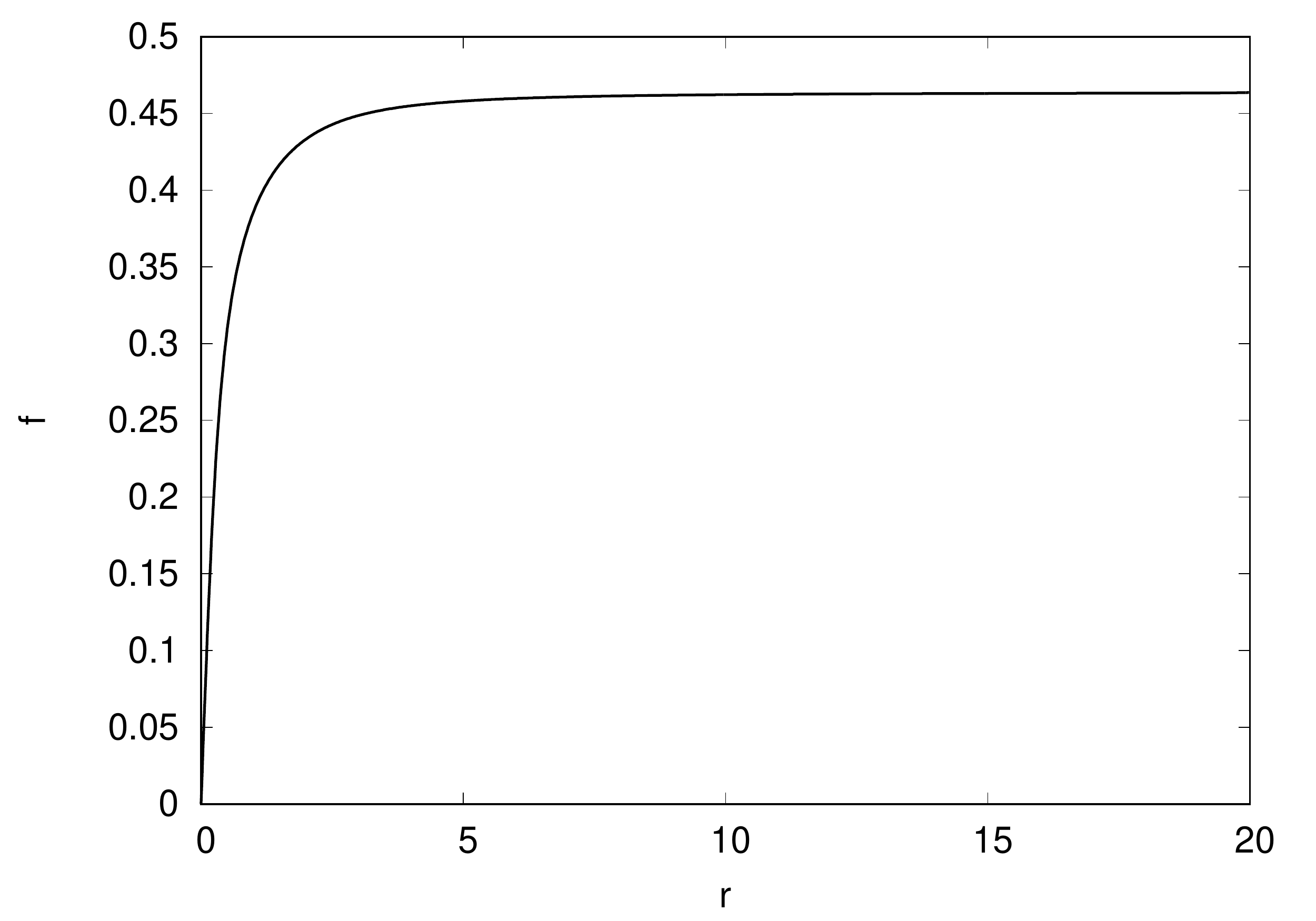}
\end{center}
\caption{Profile function $f(r)$ for the elastic constants $k_1 = k_2 = k_4 = k_5 = k_6 = 1.0$ and $k_3 = -3.0$.}
\label{profile}
\end{figure}
Fig. \ref{profile} shows the localised solution with $f(0) = 0$ corresponding to the elastic constants $k_1 = k_2 = k_4 = k_5 = k_6 = 1.0$ and $k_3 = -3.0$, which fulfil all the required constraints (\ref{elastic_constraints}) and give $\theta_0 = 0.4636$, with the parameter $\beta = 5.0$ as prescribed by analytical expressions (\ref{beta}) and (\ref{def_theta_0}). This value has been chosen since it is the optimal one in the case of the uniform distortion giving the lower energy per pitch $P=\frac{2\pi}{|\beta|}$. In fact, by varying $\beta$ we have found that still in the case of localised solutions $\beta = 5.0$ is the preferred choice, confirming that the analytical expression found for the optimal $\beta$ is still valid at least for this choice of the elastic constants.
In addition, it seems natural to stick with the analytical expression (\ref{beta}) for $\beta$ for  a better comparison with the case $\alpha=0$, as this latter reproduces the uniform heliconical state as global minimizer.
The energy per pitch of the solution corresponds to $E_{\alpha = 1}/P = -3125.97$, while for the conical distortion (which we can identify with the case $\alpha = 0$ from the general ansatz), $E_{\alpha = 0}/P = -3138.45$,
being the relative energy of the excited state $\Delta E/P = (E_{\alpha = 1} - E_{\alpha = 0})/P = 12.48$. 
As a check, the behaviour of the numerical solution around the origin has been compared with the expansion given by Eq. (\ref{symmetry_exp}) as shown in Fig. \ref{zoomup2quintic}. The value of the free parameter of the expansion $\xi$ has been taken from the numerical solution.
This result states that the localised non-uniform conical distortions are stable states with respect to the uniform nematic configuration ($\bm{n}=\bm{n}_0$) but, at the same time, they
can be seen as excitations over the ground state realised by the uniform heliconical distortion $\bm{n}=\bm{n}_h$, see equation (\ref{heliconics_1}).
\begin{figure}[ht]
\begin{center}
\includegraphics[width=.7\textwidth]{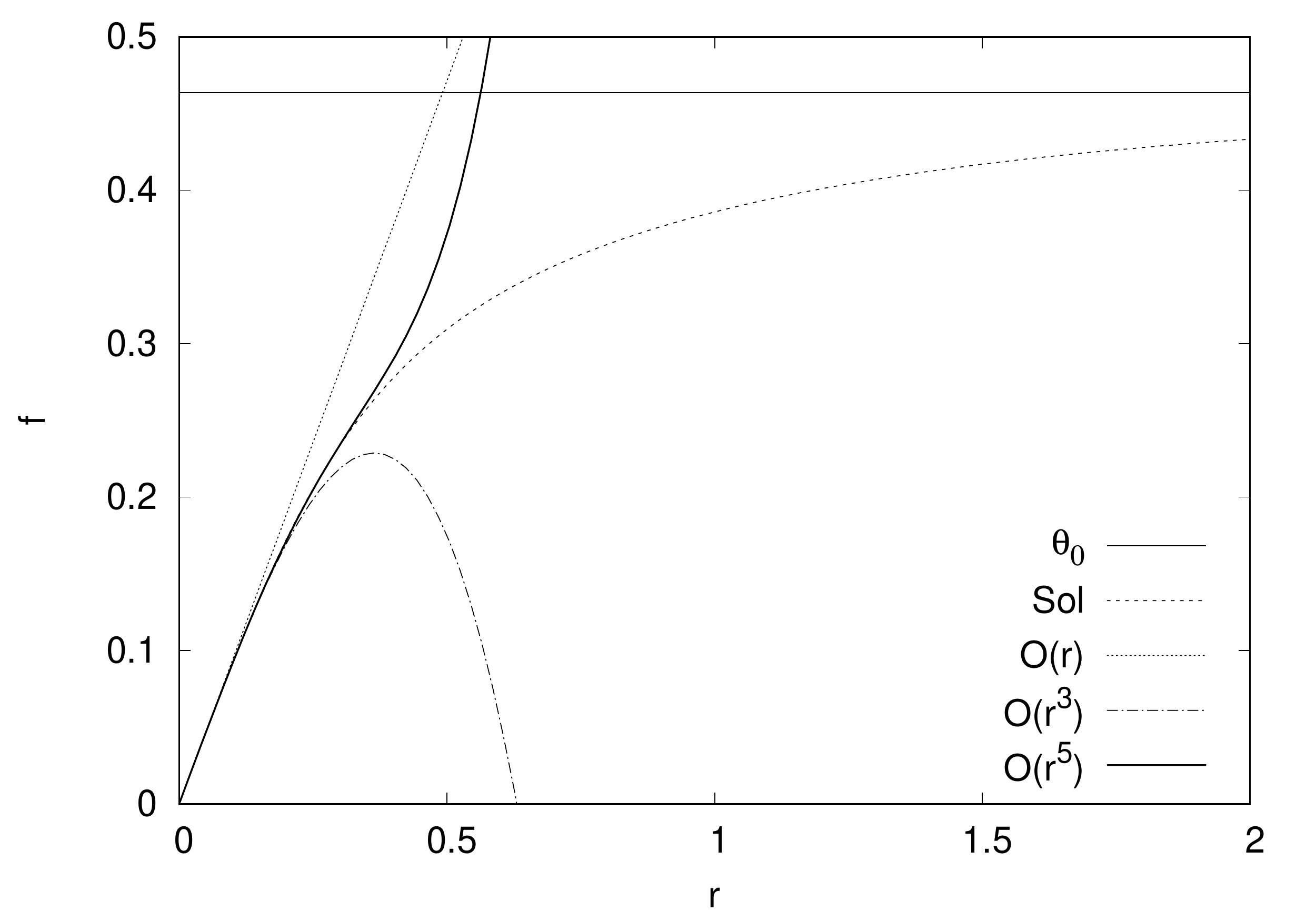}
\end{center}
\caption{Profile function $f(r)$ for $k_1 = k_2 = k_4=k_5 = k_6 = 1.0$ and $k_3 = -3.0$ together with the asymptotic value $\theta_0$ and a representation of the polynomial expansion around $r=0$ as in (\ref{symmetry_exp}) to several orders up to $r^5$ for $\xi = 0.94$.}
\label{zoomup2quintic}
\end{figure}

In Fig. \ref{3D-alpha1}, we can see the three-dimensional reconstruction of the localised configuration, where the colouring of the bars, representing the vector directors, corresponds to different values of the conical angle given by $f(r)$ as indicated. For a better understanding, this is complemented with the transversal cut on the plane $(y,z)$ appearing in Fig. \ref{alhpa1-plane&cylinder}, together with the cylindrical arrangement of those points with the same value of the conical angle.
\begin{figure}[ht] 
\begin{center}
\includegraphics[width=.8\textwidth]{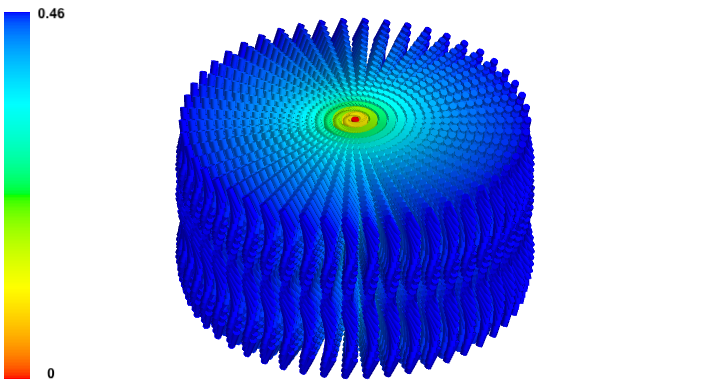}
\end{center}
\caption{Three-dimensional reconstruction of the localised solution ($\alpha = 1$) for the elastic constants $k_1 = k_2 = k_4 = k_5 = k_6 = 1.0$ and $k_3 = -3.0$. The colour bar represents the value of the conical angle $f(r)$
from the origin to the asymptotic state.}
\label{3D-alpha1}
\end{figure}
\begin{figure}[ht] 
\begin{center}
\includegraphics[width=.9\textwidth]{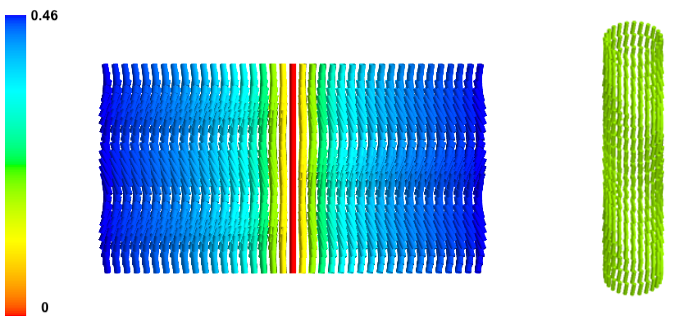}
\end{center}
\caption{Transversal cut (left) and cylinder of constant conical angle (right) for the $\alpha = 1$ case. Elastic constant values: $k_1 = k_2 = k_4 = k_5 = k_6 = 1.0$ and $k_3 = -3.0$. The colour bar represents the value of the conical angle $f(r)$
from the origin to the asymptotic state.}
\label{alhpa1-plane&cylinder}
\end{figure}

As previously commented, one can also consider the profile function taking a multiple of $\pi$ at the origin. However, since this implies a greater deviation from the conical distortion angle, the resulting configuration will have greater energy. For instance, for $f(0) = \pi$, we found that $E_{\alpha = 1}/P = -2356.40$, with a relative energy $\Delta E/P = 782.05$.

Another possibility is to study the parameter space, \emph{i. e.} the elastic constants, of the model. Note that in this case, we need to bear in mind the existing constraints (\ref{elastic_constraints}) involving them. For instance, for a fixed $k_3 = -3.0$, it is not possible to have $k_2 \geq 1.5$. Nevertheless, we can see that both $\beta$ and $\theta_0$ in (\ref{beta}) and (\ref{def_theta_0}) do not depend on the elastic constant $k_1$. Hence, it seems worth studying how the localised configurations change with an increasing value of it. However, what we find is that the energy slightly increases (see Table \ref{Table-k1}), so does the size of the configuration (see Fig. \ref{profile-k1}), although in both cases it does not seem relevant.
In particular, as for the size, it is worth noticing that the bigger $k_1$ is, the slower the asymptotic angle is approached. 
\begin{figure}[ht] 
\begin{center}
\includegraphics[width=.7\textwidth]{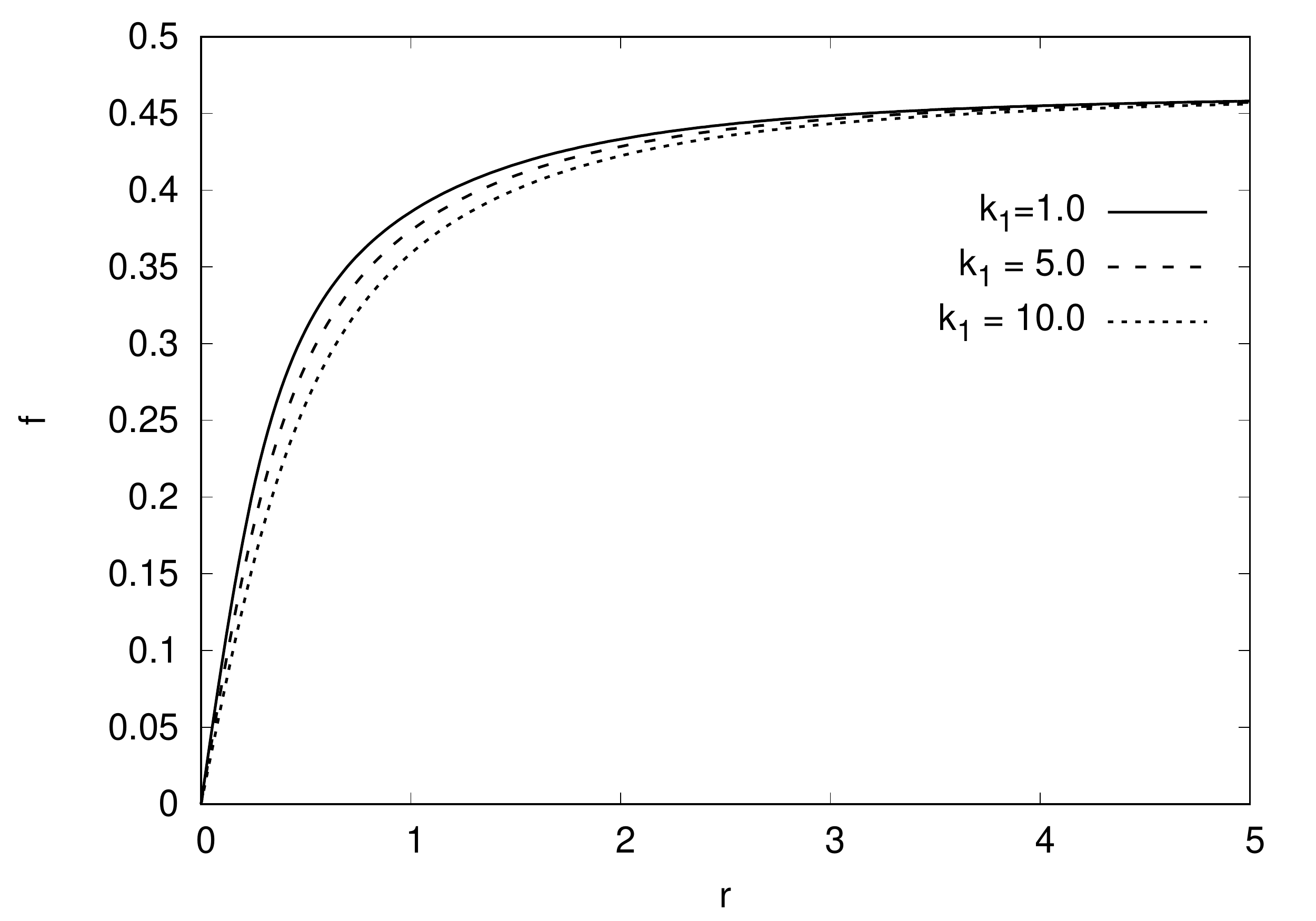}
\end{center} 
\caption{Profile function $f(r)$ near the origin for an increasing $k_1$ while $k_2 = k_4 = k_5 = k_6 = 1.0$ and $k_3 = -3.0$.
The bigger $k_1$ is, the slower the asymptotic angle is approached.}
\label{profile-k1}
\end{figure}
\begin{table}
\begin{center}
\begin{tabular}{c|c|c}
$k_1$ &  $E_{\alpha = 1}/P$ & $\Delta E/P$ \\
\hline 
\hline 
1.0 & - 3125.97 & 12.48 \\
2.0 & -3124.45 & 14.00  \\
3.0 & -3122.99 & 15.46  \\
4.0 & -3121.52 & 16.93  \\
5.0 & -3120.06 & 18.39  \\
6.0 & -3118.61 & 19.84  \\
7.0 & -3117.17 & 21.28  \\
8.0 & - 3115.73 & 22.72  \\
9.0 & -3114.31 & 24.14  \\
10.0 & -3112.89 & 25.56 
\end{tabular}
\caption{Energy per pitch and the corresponding excess as a function of $k_1$ for $k_2 = k_4 = k_5 = k_6 = 1.0$ and $k_3 = -3.0$.}
\label{Table-k1}
\end{center}
\end{table}
On the other hand, there are actually other situations where $\beta$ and $\theta_0$ change, as it can be seen from (\ref{beta}) and (\ref{def_theta_0}). For instance, this is the case when one increases the elastic constant $k_4$. As it can be seen in Table \ref{Table-k4}, this results in a decreasing of the total energy per pitch of both the uniform distortion and the localised configurations. In addition, as it can also be seen in Fig. \ref{profile-k4}, both the excitation energy $\Delta E/P$ and size of the solution are lowered, implying this latter tends to shrink for an increasing contribution of the elastic constant $k_4$.
\begin{figure}[ht]
\begin{center}
\includegraphics[width=.7\textwidth]{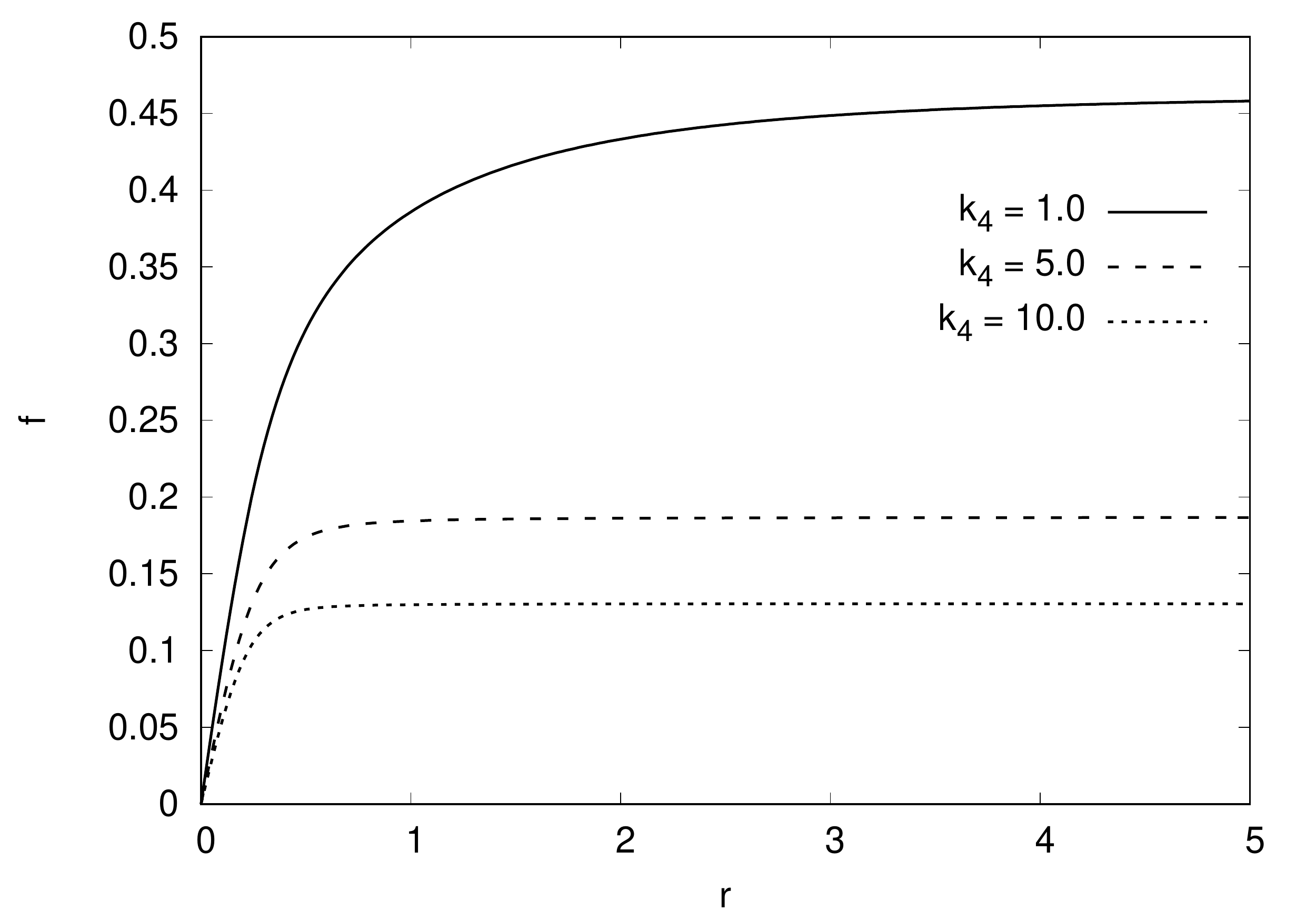}
\end{center}
\caption{Profile function $f(r)$ near the origin for an increasing $k_4$ while $k_1 = k_2 = k_5 = k_6 = 1.0$ and $k_3 = -3.0$.
The bigger $k_4$ is, the faster the asymptotic angle is approached.}
\label{profile-k4}
\end{figure}
\begin{table}
\begin{center}
\begin{tabular}{c|c|c|c|c|c}
$k_4$ & $\beta$ & $\theta_0$ &  $E_{\alpha = 0}/P$ & $E_{\alpha = 1}/P$ & $\Delta E/P$ \\
\hline
\hline
1.0 & 5.0 & 0.4636 & -3138.45 & -3125.97 & 12.48 \\
2.0 & 6.3509 & 0.3063 & -2929.22 & -2924.48 & 4.74 \\
3.0 & 7.6026 & 0.2450 & -2887.37 & -2884.37 & 3.00 \\
4.0 & 8.6932 & 0.2101 & -2869.44 & -2867.20 & 2.24 \\
5.0 & 9.6667 & 0.1868 & -2859.48 & -2857.67 & 1.81 \\
6.0 & 10.5529 & 0.1698 & -2853.14 & -2851.61 & 1.53 \\
7.0 & 11.3714 & 0.1568 & -2848.75 & -2847.42 & 1.33 \\
8.0 & 12.1353 & 0.1464 & -2845.53 & -2844.35 & 1.18 \\
9.0 & 12.8544 & 0.1378 & -2843.07 & -2842.00 & 1.07 \\
10.0 & 13.5355 & 0.1306 & -2841.12 & -2840.15 & 0.97
\end{tabular}
\caption{Energy per pitch (both of the uniform distortion and localised configuration) and the corresponding excess $\Delta E/P$ as a function of $k_4$ for $k_1 = k_2 = k_5 = k_6 = 1.0$ and $k_3 = -3.0$.}
\label{Table-k4}
\end{center}
\end{table}
Finally, we can also easily study the behaviour of the localised solution when decreasing $k_3$ from $-3$ to $-10$ (see Fig. \ref{profile-k3} and Table \ref{Table-k3}). In this case, the size also decreases accompanied by an increasing in the excitation energy.
\begin{figure}[ht]
\begin{center}
\includegraphics[width=.7\textwidth]{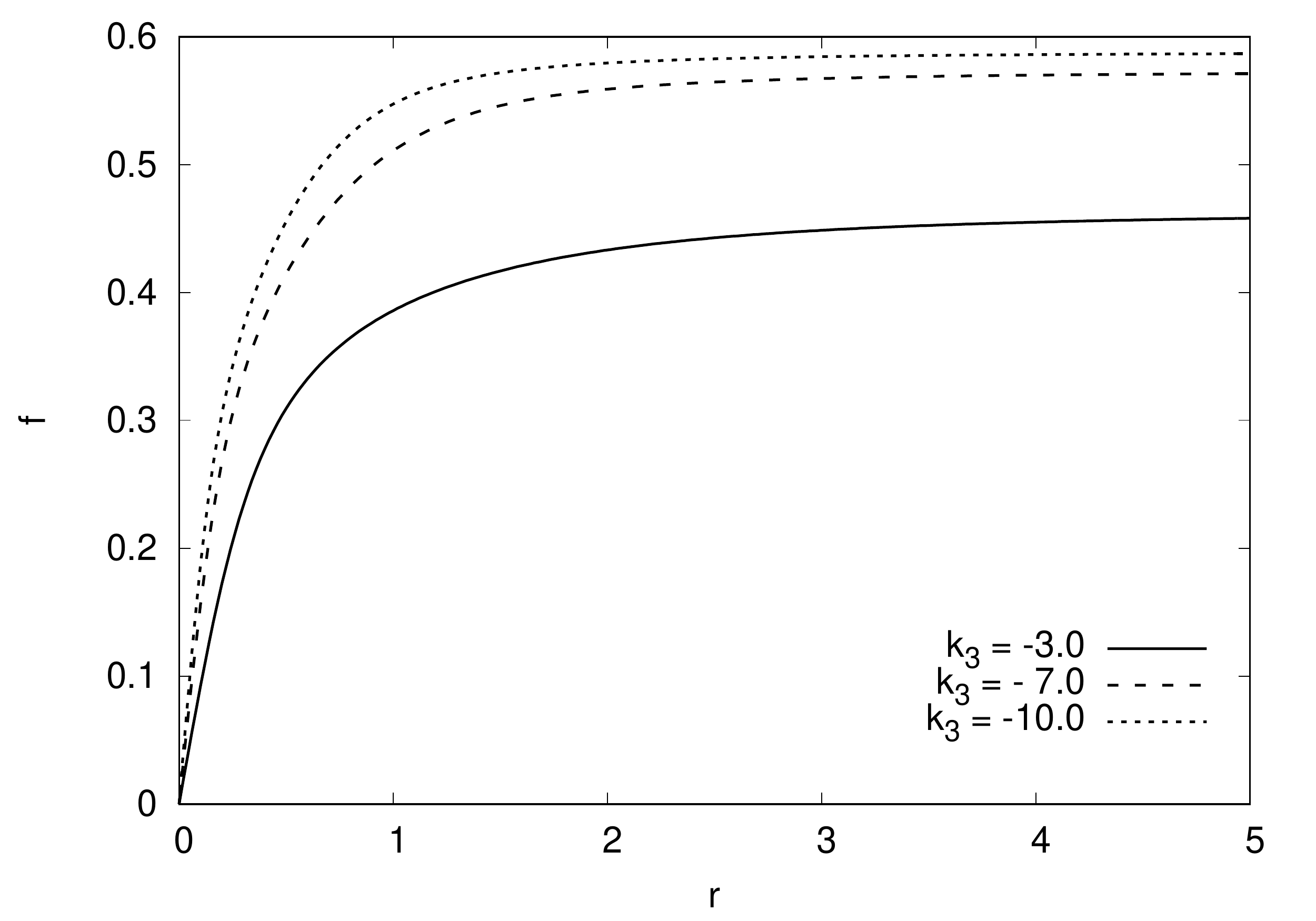}
\end{center}
\caption{Profile function $f(r)$ near the origin for an increasing $k_3$ while $k_1 = k_2 = k_4 = k_5 = k_6 = 1.0$.
The bigger $|k_3|$ is, the faster the asymptotic angle is approached.}
\label{profile-k3}
\end{figure}
\begin{table}
\begin{center}
\begin{tabular}{c|c|c|c|c|c}
$k_3$ & $\beta$ & $\theta_0$ &  $E_{\alpha = 0}/P$ & $E_{\alpha = 1}/P$ & $\Delta E/P$ \\
\hline
\hline
-3.0 & 5.0 & 0.4636 & -3138.45 & -3125.97 & 12.48 \\
-4.0 & 5.6569 & 0.5236 & -6276.90 & -6250.90 & 26.00 \\
-5.0 & 6.3509 & 0.5495 & -10670.7 & -10629.9 & 40.8 \\
-6.0 & 7.0 & 0.5639 & -16319.9 & -16263.6 & 56.3 \\
-7.0 & 7.6026 & 0.5732 & -23224.5 & -23152.0 & 72.5\\
-8.0 & 8.1650 & 0.5796 & -31384.5 & -31295.4 & 89.1 \\
-9.0 & 8.6932 & 0.5844 & -40799.9 & -40693.8 & 106.1 \\
-10.0 & 9.1924 & 0.5880 & -51470.6 & -51347.2 & 123.4
\end{tabular}
\caption{Energy per pitch (both of the uniform distortion and localised configuration) and the corresponding excess as a function of $k_3$ for $k_1 = k_2 = k_4 = k_5 = k_6 = 1.0$.}
\label{Table-k3}
\end{center}
\end{table}
The above numerical results for all the analysed cases have also been confirmed by using a shooting method for equation \eqref{eq:ode1}, together with the use of an adaptive mesh in order to cope with the stiffness of the equation around the origin.
For this alternative numerical method the normal form of equation (\ref{eq:ode1}) has been used as reported in \cite{SupMat}.

Now that we have studied the solutions to equation \eqref{eq:ode1} by numerical methods, we can check the goodness of approximation \eqref{fapp4} by a best fitting procedure. As mentioned above, the number of free parameters for least square minimization can be reduced to two, \emph{i. e.} $(\xi, b_2)$, by means of \eqref{104_eq_f0} and \eqref{rela_first_third}. 
However, here first we use all four parameters $(\xi,\zeta,b_2,f_0)$ for a few examples.
Then we provide the results of the fitting procedure leaving free  only ($\xi, b_2$) or just $b_2$ (with $\xi$ fixed by numerics), for the case $k_1 = k_2 = k_4 = k_5 = k_6 = 1.0,\; k_3 = -3.0$ shown in Fig. \ref{profile}.
By doing so, we show the remarkable capability of \eqref{fapp4} to adapt itself to the numerical solutions.

The results of the procedure using all the four possible parameters are provided in table \ref{table:fit} for four different sets $\lbrace k_{i}\rbrace$. Here, the values of the best fitting parameters together with the reference values $\xi_{\text{num}}$, $\zeta_{\text{num}}$ extrapolated
from numerical solutions are provided.
As an estimator of the goodness of the best fit, we report in the last column the distance $||f-f_a||_2$ between the numerical solution $f$ and the approximation $f_a$, this latter obtained by replacing in (\ref{fapp4}) the parameters $(\xi, \zeta, b_2, f_0) $ with the best fitted ones.

Moreover, the latter results are depicted in Fig. \ref{fig:fit2}, in order to provide a visual representation of them. Finally, the detailed analysis, with two and one free parameters respectively, for the case $k_1=k_2=k_4=k_5=k_6=1$, $k_3=-3$ is displayed in Fig. \ref{fig:fit}. 

According to these results, we can conclude that \eqref{fapp4} is a quite good approximation for the solutions of \eqref{eq:ode1}.
  
 \begin{center}  
 \begin{table}
 \begin{tabular}{c|c|c|c|c|c|c|c}
  Case &  $\xi$ & $\xi_{\text{num}}$ & $\zeta$ & $\zeta_{\text{num}}$ & $f_0$ & $b_2$ & $||f-f_a||_2$\\  
  \hline 
    1 & 0.9620 & 0.9465 & -2.8997  & -2.3733 &  0.4637   &  3.9775      & 0.0063\\
    2 & 0.6604 & 0.6650 &  -2.1686 & -2.5290 & 0.1868   & 95.2940       & 0.0015           \\
    3 & 0.6854 & 0.6793 &  -1.0512 & -0.9496 &  0.4637   &  1.9905      & 0.0077           \\
    4 & 1.4608 & 1.4400 & -13.8035 & -11.4797& 0.3368   & 41.6260       & 0.0024       
\end{tabular}
\caption{\label{table:fit}   Results of the best fitting procedure for four different sets of $\lbrace k_i\rbrace$: $k_1=k_2=k_4=k_5=k_6=1,\; k_3=-3$ (Case 1); $k_1=k_2=k_5=k_6=1,\; k_3=-3,\; k_4=5$ (Case 2); $k_1=10,\; k_2=k_4=k_5=k_6=1,\; k_3=-3$ (Case 3); $k_1=k_4=k_5=k_6=1,\; k_2=3,\; k_3=-7$ (Case 4). The quantities $\xi_{\text{num}}$ and $\zeta_{\text{num}}$ represent the values obtained from numerical solutions, while $f_a$ is the approximation obtained by replacing in (\ref{fapp4}) the parameters $(\xi, \zeta, b_2, f_0) $ with the best fitted ones. $f_0$ is obtained from the best fitted and it reproduces up to the fourth decimal digit the value from (\ref{104_eq_f0}).}
\end{table}
\end{center} 
\begin{figure}[t]
\centering
\subfigure{\includegraphics[width=.45\textwidth]{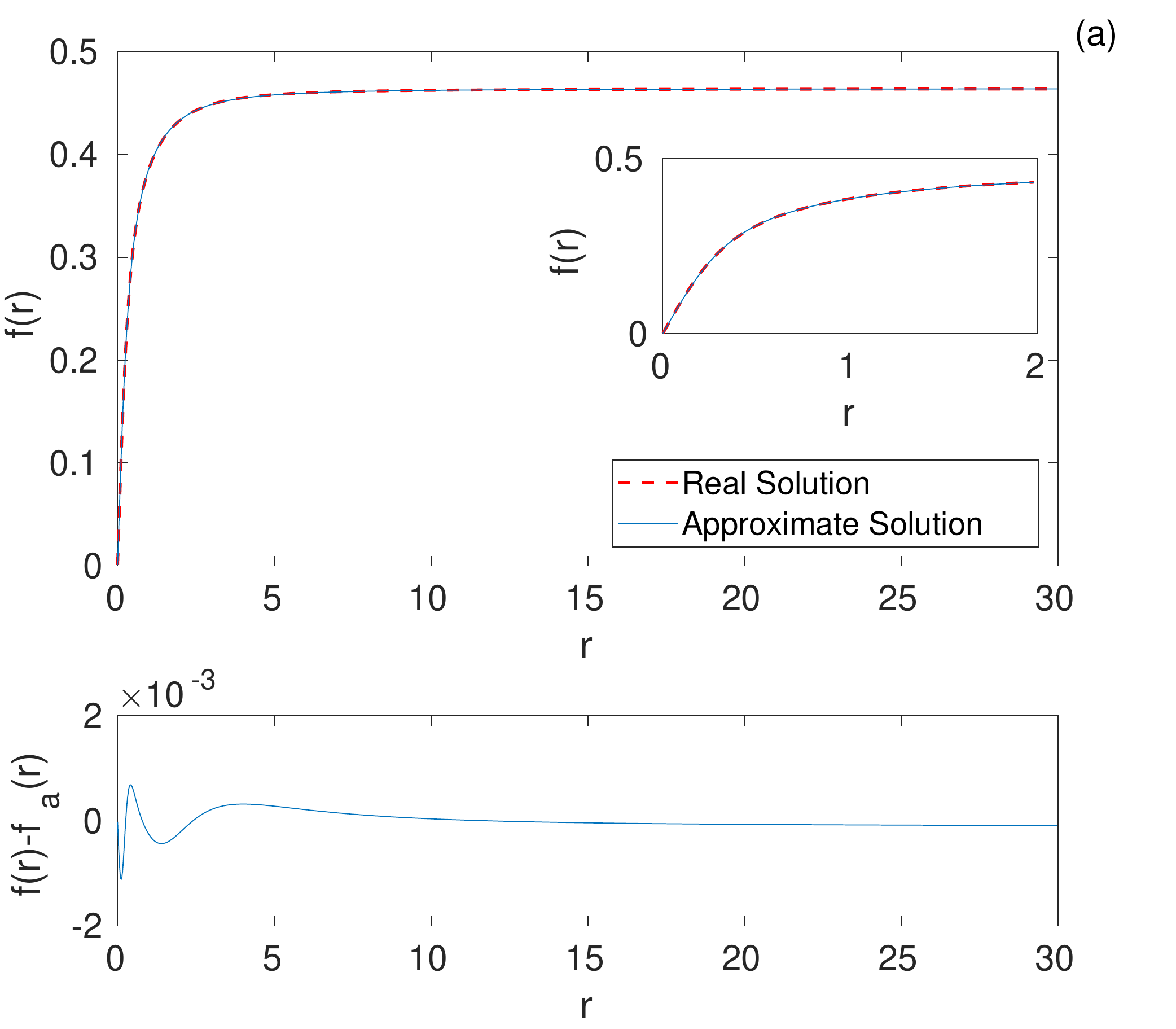}}
\subfigure{\includegraphics[width=.45\textwidth]{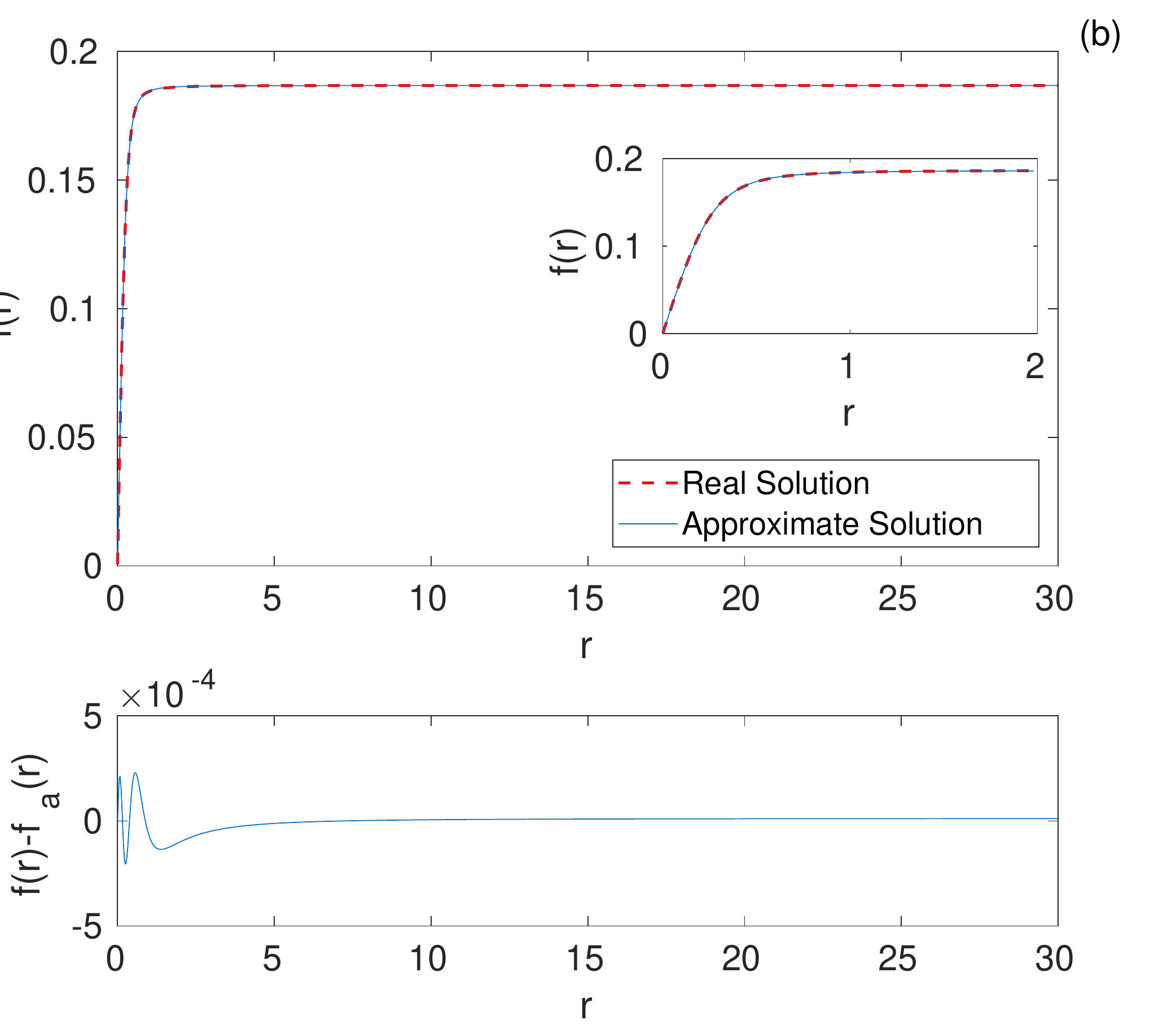}}
\subfigure{\includegraphics[width=.45\textwidth]{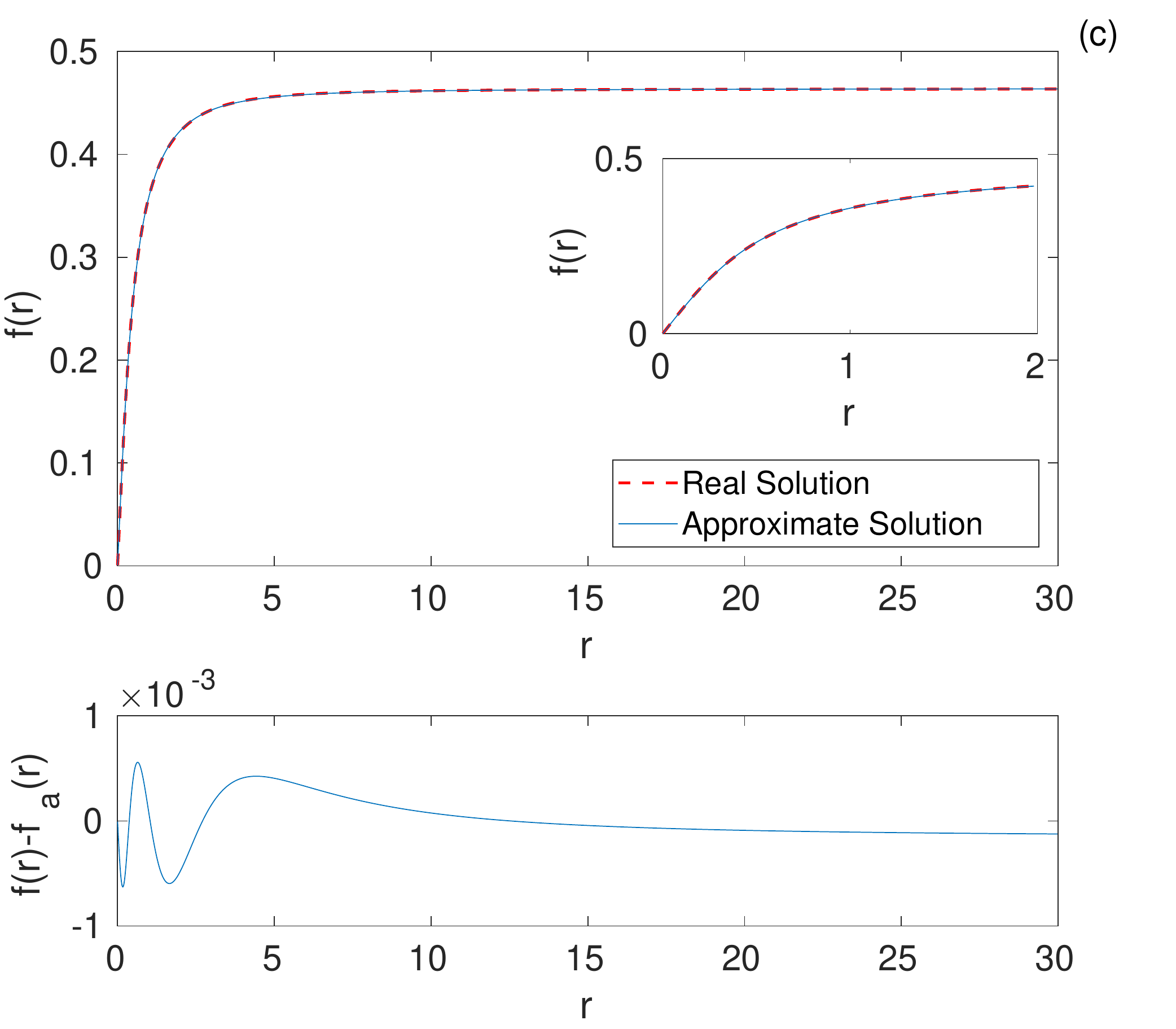}}
\subfigure{\includegraphics[width=.45\textwidth]{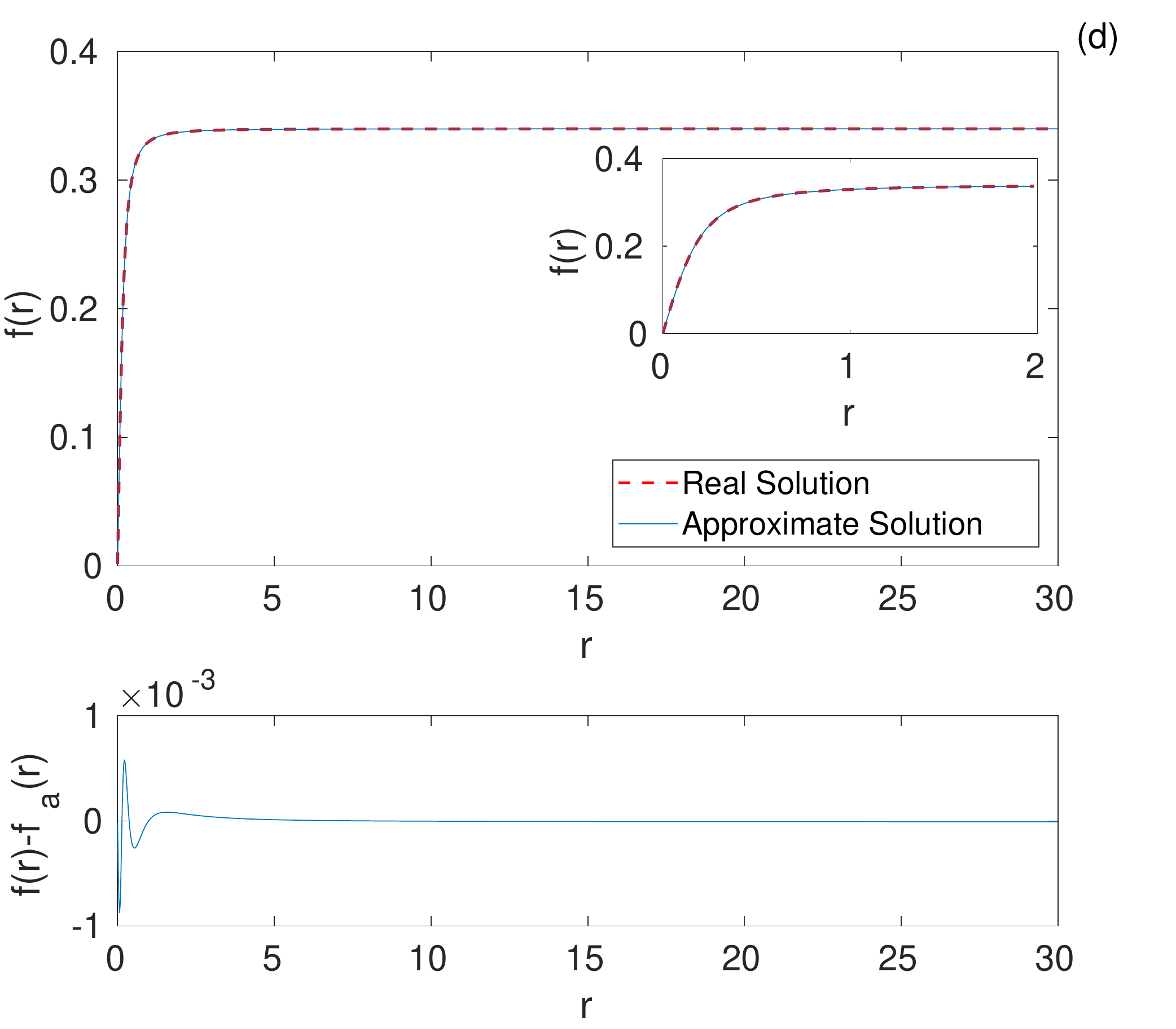}}
\caption{Best fits between the numerical solution and approximation \eqref{fapp4} corresponding to the cases listed in Table \ref{table:fit}: (a) $k_1=k_2=k_4=k_5=k_6=1,\; k_3=-3$; (b) $k_1=k_2=k_5=k_6=1,\; k_3=-3,\; k_4=5$;
(c) $k_1=10,\; k_2=k_4=k_5=k_6=1,\; k_3=-3$; (d) $k_1=k_4=k_5=k_6=1,\; k_2=3,\; k_3=-7$.}
\label{fig:fit2}
\end{figure} 
\begin{figure}[t] 
\centering
\subfigure{\includegraphics[width=.45\textwidth]{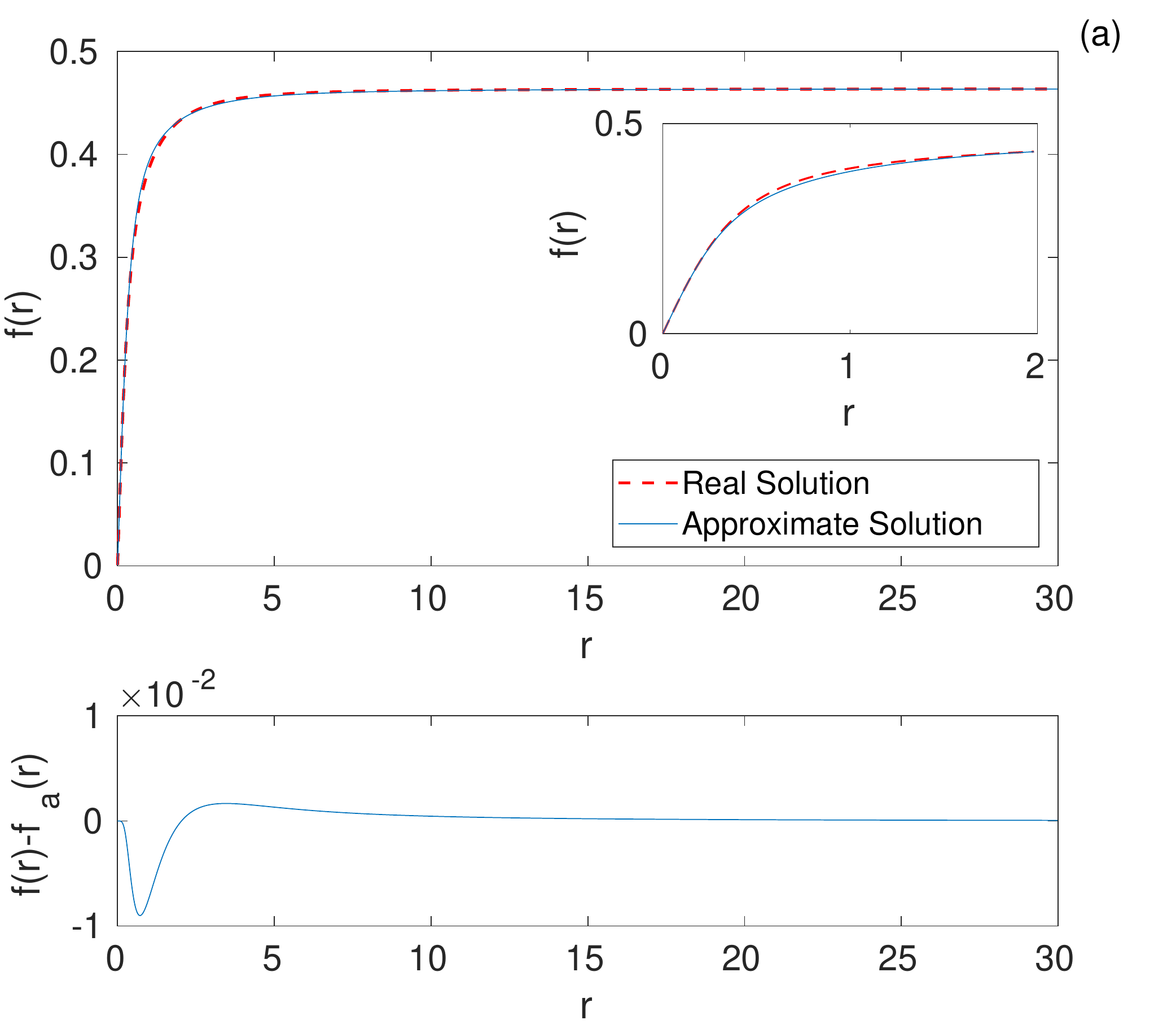}}
\subfigure{\includegraphics[width=.45\textwidth]{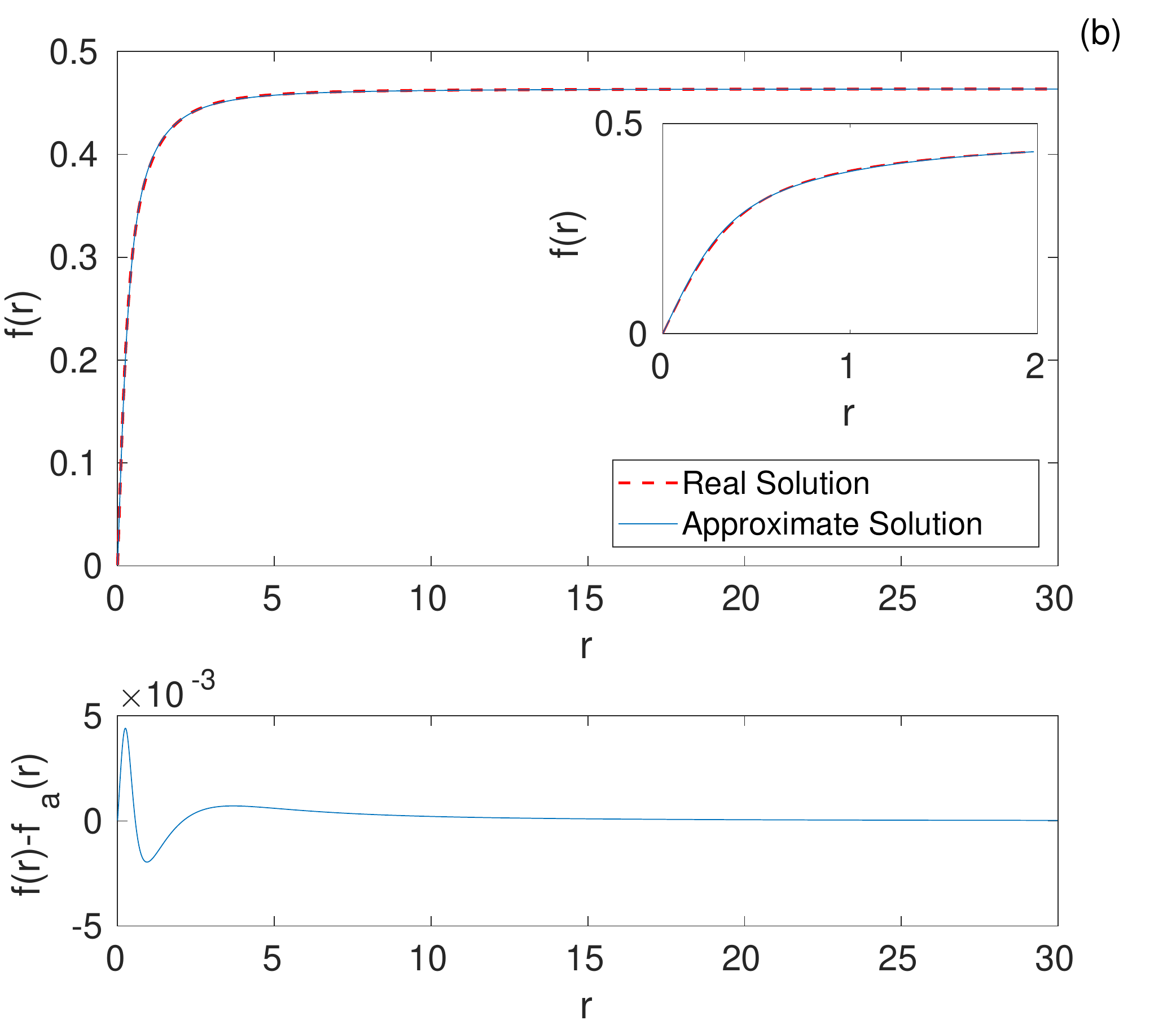}}     
\caption{Best fits between the numerical solution and approximation  \eqref{fapp4} leaving free (a) $b_2$ and (b) both $\xi,\; b_2$, for the case $k_1=k_2=k_4=k_5=k_6=1$, $k_3=-3$. Here (a) $b_2=1.4884$ and (b) $(b_2,\;\xi)=(2.5594
,\;0.9217)$.}
\label{fig:fit}    
\end{figure}
\begin{figure}[ht] 
\begin{center}
\includegraphics[width=.45\textwidth]{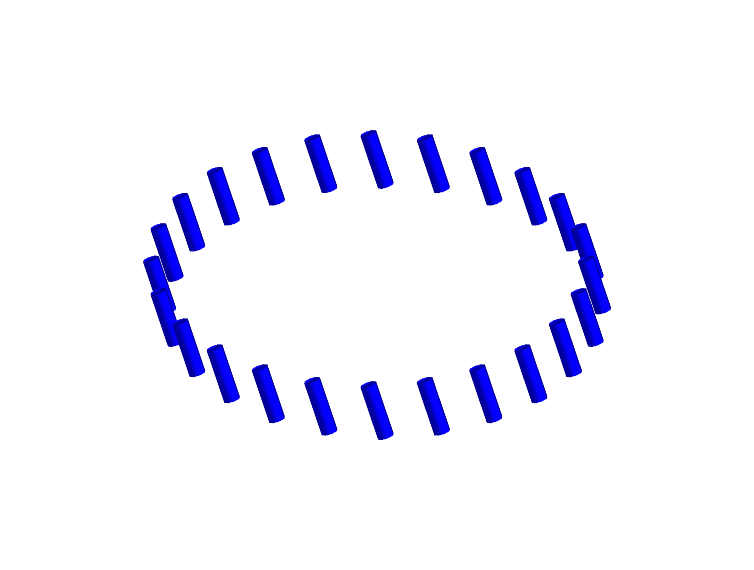}
\includegraphics[width=.45\textwidth]{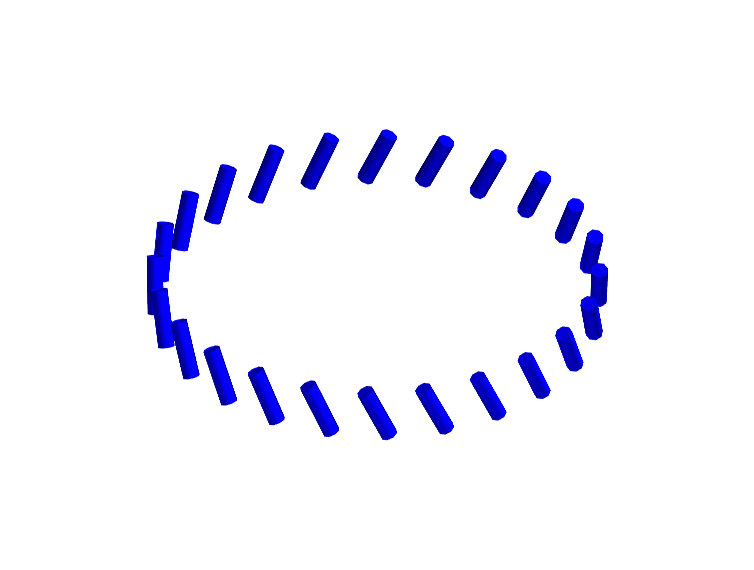}
\end{center}
\caption{Circle of constant radius for the solutions of $\alpha = 0$ (left) and $\alpha=1$ (right) far from the origin at fixed $z$. Elastic constant values: $k_1 = k_2 = k_4 = k_5 = k_6 = 1.0$ and $k_3 = -3.0$.}
\label{Circles}
\end{figure}

\section{Conclusions and Perspectives}
\label{sectfive}

In this paper we studied a generalised elasticity theory for liquid crystals put forward recently in \cite{Virga4}, parameterised by six elastic constants: $k_1$, $k_2$, $k_3$ coming from the standard Frank energy second order contributions and $k_4$, $k_5$, $k_6$ related to the fourth order terms, as shown in Eq. (\ref{virga_free_energy}) and constrained by Eq. (\ref{elastic_constraints}). The quartic part, positive definite, is needed to stabilize the negative contribution from $k_3 < 0$ for the heliconical phase to take place. Thus, the proposed theory generalizes Frank's elastic energy density
to include quartic terms in the spatial gradients of the nematic director. For a suitable choice of the elastic constants, the novel free-energy functional admits heliconical configurations as ground state.
These ground states have been determined by minimising the free-energy density with respect to the two parameters, $\beta$ and $\theta_0$, of the heliconical solution.
In the present paper we have adopted a different approach, using the Euler-Lagrange equations. We have determined the pitch and the conical angle of the
heliconical configurations. After that, we generalised the heliconical configurations to non-uniform structures with a variable conical angle (\ref{general_ansatz}).
The generalised solution contains two parameters $\alpha, \beta$ and a profile function for the conical angle depending on the radial distance from the symmetry axis of the
configuration.
We have studied the Euler-Lagrange equation associated with the reduced functional on this family of solutions in two distinct cases: $\alpha=0$ and $\alpha=1$. This $\alpha$ parameter describes nothing but how the vector director winds around the $z$-axis (see Fig. \ref{Circles}). When it vanishes there is no winding, while it goes around the $z$-axis once when $\alpha=1$. These structures may resemble to vortex-like configurations with $\alpha$ taking the role of the vortex charge.
We have performed both numerical and analytical studies and we found a non-uniform profile function only in the case $\alpha=1$. Case $\alpha=0$ corresponds
to the uniform heliconical solutions found in \cite{Virga4}, 
its structure is a pile of different strata, each of them with the same constant conical angle $\theta_0$, as shown in Fig. \ref{3D-alhpa0}, continuously precessing as moving parallel to the $z$-axis. In contrast, when $\alpha = 1$, there is a simultaneous bending of the conical angle, from 0 to $\theta_0 $ in the radial direction, which is precessing both in the $z$-direction and by azimuthal rotations. Thus, an helix appears as a line for each fixed constant director, together with the already mentioned winding around the $z$-axis. The corresponding energy still remains under the uniform nematic  configuration. However, the system needs to go through an unwinding before reaching the uniform heliconical distortion, and in this sense these solutions can be seen as stable excitations of them.

It is worth comparing our results with the work presented in \cite{PRB100,PRB98}. There, similar configurations called Skyrmion tubes have been numerically described in ferromagnets, while experimentally found in liquid crystals. However, although similarities between ferromagnets and liquid crystals are well-known, there was no theoretical description for this kind of configurations in achiral nematics in the absence of external fields . Then, at least at the best of our knowledge, the quartic degree free-energy proposed by Virga and studied here is the first theory within liquid crystals supporting these localised structures.

Indeed, our solutions described above are quite similar to those in \cite{PRB100,PRB98}, with the main difference being the asymptotic behaviour far from the origin. As opposed to their case, where the vector director achieves the uniform distortion state (the case $\alpha = 0$ in our language), ours is given by a vector director with a constant conical angle, but presenting a winding around the symmetry axis as well. Hence, the configurations studied here might provide a good description of the cores of the Skyrmion tubes. 
The similiraties between our configurations and those found in cholesterics should not surprise. In chiral ferromagnets and liquid crystals the lack of inversion symmetry, due to the presence of the antisymmetric contributions like  Dzyaloshinskii-Moriya interaction or $q_0 \, \boldsymbol{n} \cdot {\rm curl} \boldsymbol{n}$, together with the frustration from geometric confinement and external fields, creates competing effects which may lead to the formation of a vast class of defects in the director distribution, either topological or not. Although in our case the inversion symmetry still holds, a similar mechanism of competition between the quadratic part of the functional with negative $k_3$ and the positive definite quartic part leads to non-trivial localised configurations like those we found, even in the absence of external frustration.
On the other hand, it should be noted that we have shown how, far from the origin, the dominant contribution to the free-energy is exactly the same both for $\alpha = 0$ and $\alpha=1$ cases so, at least at that level, they are equivalent.
Thus, despite the cylindrical symmetry of our ansatz prevents us from joining the winding localised solution with $\alpha = 1$ to the uniform distortion, this seems to indicate that the skyrmionic structures in \cite{PRB100,PRB98} may also be supported by this quartic theory.
Although it is outside the scope of this paper,  a more general ansatz will be pursued in the future.

In this paper, we have shown how the first non-zero higher (quartic) order terms in a theory with negative $k_3$ can host localized distortions in achiral nematics, while Frank-Oseen quadratic free-energy functional allows them only in the chiral case under external fields. This suggests that structures resembling our solutions might be experimentally found without applying external fields.

On the other hand, the expression given  in (\ref{fapp4}) opens new possibilities in the study of field equations of interest, like (\ref{eq:ode1}), in the domain of Skyrmions and similar configurations in liquid crystals and  magnetic materials. Generally, they are only addressed by numerical methods, because of the  complicated structure of different effects at different scales.
In our particular case, these effects are related to the singularities in the coefficients at the origin  and the appearance of the trigonometric multiple field contributions in the free-energy.
The procedure leading to (\ref{fapp4}) is based on
a systematic and algorithmic manipulation of analytic expressions which closely reproduce the numerical solutions, even if they do not provide the exact results.
In order to obtain this, the method of the Pad\'e approximants has played an important role. Actually, by our mixed method we proved that the 4th order approximant   already provides an accuracy of  $10^{-3}$ 
in reproducing the numerical solution, by a suitable choice of the parameters.

In the future, we would like to develop and apply a coherent procedure leading to an accurate {\emph{a priori}} evaluation of the Pad\'e coefficients at a given order of approximation.
The study of the singularities  of the approximated solution  in the complex $r$-plane may indicate how to tackle such a problem in an efficient way, {\emph{e. g.}} moving or adding poles  on suitable conjugated points. 

In addition, we also plan to study the interactions of the obtained localised structures among them and the effect of the interaction with external electric/magnetic fields in order to control the main structure parameters.
We also aim at studying the proposed quartic free-energy functional in confined geometries for liquid crystals.

\section{Acknowledgments}   
GDM is supported by the Dipartimento di Matematica e Fisica "E. De Giorgi", University of Salento grant {\it Studio analitico di configurazioni spazialmente localizzate in materia condensata e materia nucleare}.
CN is supported by the INFN grant  19292/2017 (MMNLP) {\it Integrable Models and Their Applications to Classical and Quantum Problems}.  LM has been partially supported by  INFN IS-MMNLP. VT is partially supported
by Italian Ministry of Education, University and Research MIUR.  

\appendix

\section{Mathematical details}
\label{app:equation}

In this Appendix we collect the basic main functions and coefficients appearing in the equilibrium equations for both cases $\alpha=0$ and $\alpha=1$.

\subsection{Case: $\alpha=0$}

The quantities $\Gamma_i(f)$, with $i=0,2,4$, involved in the reduced free-energy for $\alpha = 0$ (see Eqs. (\ref{free_energy_alphazerozero}) and (\ref{alphazeroode}) in the main text) are given by
\begin{equation}
\Gamma_0(f) = \gamma_{01} + \gamma_{02} \cos(2f)+\gamma_{03} \cos(4f) + \gamma_{04} \cos(6f) + \gamma_{05} \cos(8f),
\end{equation}
with coefficients
\begin{eqnarray}
\gamma_{01} &=&\frac{\pi^2}{128} \beta [32 ( 6 k_2 + k_3) + (70 k_4 + 3 k_5 - 10k_6) \beta^2], \\
\gamma_{02} &=& - \frac{\pi^2}{16} \beta [32 k_2 + (14 k_4 - k_6)\beta^2], \\
\gamma_{03} &=& \frac{\pi^2}{32} \beta [16 k_2 - 8 k_3 + (14 k_4 - k_5 + 2 k_6) \beta^2], \\
\gamma_{04} &=& - \frac{\pi^2}{16} (2 k_4 + k_6) \beta^3,\\
\gamma_{05} &=& \frac{\pi^2}{128} (2 k_4 + k_5 + 2 k_6) \beta^3.
\end{eqnarray}
Then
\begin{eqnarray}
\Gamma_2(f) &=& \gamma_{21} + \gamma_{22} \cos 2 f + \gamma_{23} \cos 4 f  + \gamma_{24} \cos 6 f,
\end{eqnarray}
with
\begin{eqnarray}
\gamma_{21} &=& \frac{\pi^2}{32\beta}\left(16k_1+80k_2+16k_3+74\beta^2k_4+2\beta^2k_5+18\beta^2k_6\right), \\
\gamma_{22} &=& \frac{\pi^2}{32 \beta}\left(16k_1+16k_2-16k_3-97\beta^2k_4-\beta^2k_5-17\beta^2k_6\right), \\
\gamma_{23} &=& \frac{\pi^2}{32}\beta\left(22k_4-2k_5-2k_6\right), \\
\gamma_{24} &=& \frac{\pi^2}{32}\beta\left(k_4+k_5+k_6\right).
\end{eqnarray}
And finally,
\begin{equation}
\Gamma_4(f) = \gamma_{41} + \gamma_{42} \cos(2f) + \gamma_{43} \cos(4f),
\end{equation}
with
\begin{eqnarray}
\gamma_{41} &=& \frac{\pi^2}{64\beta}\left(65k_4+9k_5-8k_6\right), \\
\gamma_{42} &=& \frac{\pi^2}{64\beta}\left(20k_4-12k_5+8k_6\right), \\
\gamma_{43} &=& \frac{\pi^2}{64\beta}\left(3k_4+3k_5\right).
\end{eqnarray}

\subsection{Case: $\alpha=1$}

The quantities $G_i$, $i = 0, 1, 2, 3, 4$ appearing in equation (\ref{eq:ode1}) depend on $r, f, \beta, k_1, k_2, k_3, k_4, k_5, k_6$ and are listed below:
\begin{equation}
G_0 = G_0\left(r, f\right)=
g_{01}+g_{02}\cos(2f)+g_{03}\cos(4f)+g_{04}\cos(6f)+g_{05}\cos(8f) ,
\end{equation}
where
\begin{eqnarray}
g_{01}& =& \frac{1}{16r^3}(178k_4+105k_5-30k_6)+\frac{1}{2r}(64k_1+96k_2+48k_3+70\beta^2k_4+15\beta^2k_5-18\beta^2k_6)\nonumber\\
&+&\frac{\beta^2}{2}r(192k_2+32k_3+70\beta^2k_4+3\beta^2k_5-10\beta^2k_6),
\end{eqnarray}
\begin{eqnarray}
g_{02}&=&-\frac{1}{2r^3}(25k_4+21k_5-3k_6)-\frac{32}{r}(k_1+k_2+k_3)-\frac{2\beta^2}{r}(21k_4+3k_5-10k_6)\nonumber\\
&-&4\beta^2 r(32k_2+14\beta^2k_4-k_6\beta^2),
\end{eqnarray}
\begin{eqnarray}
g_{03}&=&\frac{1}{4r^3}(2k_4+21k_5+6k_6)-\frac{2}{r}(8k_2-4k_3+3\beta^2k_5+10\beta^2k_6)\nonumber\\
&+&2\beta^2 r(16k_2-8k_3+14\beta^2 k_4-\beta^2 k_5+2\beta^2 k_6),
\end{eqnarray}
\begin{eqnarray}
g_{04}&=&-\frac{1}{2r^3}(-k_4+3k_5+3k_6)+\frac{2\beta^2}{r}(5k_4+3k_5+6k_6)-4\beta^4 r (2k_4+k_6),
\end{eqnarray}
\begin{eqnarray}
g_{05}&=&(2k_4+k_5+2k_6)(\frac{3}{16r^3}-\frac{3\beta^2}{2r}+\frac{\beta^4}{2}r),
\end{eqnarray}
with
\begin{equation}
g_{01}+g_{02}+g_{03}+g_{04}+g_{05}=0.
\end{equation}
As for $G_1$
\begin{eqnarray}
G_1=G_1(r,f)=g_{11}\sin{(2f)}+g_{12}\sin(4f)+g_{13}\sin(6f),
\end{eqnarray}
where
\begin{eqnarray}
g_{11} = \frac{1}{r^2}(-35k_4+5k_6)+64(k_1-k_2)-20(k_4+k_6)\beta^2,
\end{eqnarray}
\begin{eqnarray}
g_{12} = \frac{4}{r^2}(4k_4-k_6)+16(k_4+k_6)\beta^2,
\end{eqnarray}
\begin{eqnarray}
g_{13} = \frac{1}{r^2}(k_4+k_6)-4(k_4+k_6)\beta^2.
\end{eqnarray}
As for $G_2$
\begin{eqnarray}
G_2=G_2(r,f)=g_{21}+g_{22}\cos(2f)+g_{23}\cos(4f)+g_{24}\cos(6f),
\end{eqnarray}
where
\begin{eqnarray}
g_{21}=\frac{1}{r}(71k_4+5k_5+29k_6)+4r\left[8(k_1+5k_2+k_3)+\beta^2(37k_4+k_5+9k_6)\right],
\end{eqnarray}
\begin{eqnarray}
g_{22} = -\frac{1}{2r}(15k_4+15k_5+79k_6)+2r\left[16(k_1+k_2-k_3)-\beta^2(97k_4+k_5+17k_6)\right],
\end{eqnarray}
\begin{eqnarray}
g_{23}=\frac{1}{r}(-63k_4+3k_5+11k_6)+4\beta^2 r(11k_4-k_5-k_6),
\end{eqnarray}
\begin{eqnarray}
g_{24}=(k_4+k_5+k_6)(2\beta^2 r-\frac{1}{2r}).
\end{eqnarray}
The function $G_3$ is given by
\begin{equation}
G_3 = G_3(f) =g_{31}\sin(2f)+g_{32}\sin(4f),
\end{equation}
where
\begin{equation}
g_{31}=-8(6k_4+k_6),
\end{equation}
\begin{equation}
g_{32}=-4(4k_4-k_6).
\end{equation}
Finally,
\begin{equation}
G_4 = G_4(r,f)=g_{41}+g_{42}\cos(2f)+g_{43}\cos(4f),
\end{equation}
where
\begin{eqnarray}
g_{41} = r(65k_4+9k_5-8k_6),
\end{eqnarray}
\begin{eqnarray}
g_{42}=4r(5k_4-3k_5+2k_6),
\end{eqnarray}
\begin{eqnarray}
g_{43}=3r(k4+k5).
\end{eqnarray}



\newpage

\setcounter{section}{19}
\section*{Supplemental Material}

\setcounter{subsection}{0}
\setcounter{equation}{0}

\subsection{Uniform heliconical distortions}

In the following we will show that the heliconical configurations
\begin{equation}
\label{heliconics_1}
\bm{n}_h=\sin\theta_0\cos\beta z\bm{e}_x+\sin\theta_0\sin\beta z\bm{e}_y+\cos\theta_0\bm{e}_z,
\end{equation}
are actually solutions of the Euler-Lagrange equations associated to the functional 
\begin{eqnarray}
\label{virga_free_energy2}
F_{TB}&=&\frac{1}{2}(k_1-k_2)(\text{div}\bm{n})^2+k_2(\bm{n}\cdot\text{curl}\bm{n})^2+k_2\text{tr}(\nabla\bm{n})^2+\frac{1}{2}k_3|\bm{n}\times\text{curl}\bm{n}|^2+
\frac{1}{4}k_4(\bm{n}\cdot\text{curl}\bm{n})^4\nonumber\\
&+&
k_4\left[\text{tr}(\nabla\bm{n})^2
+\frac{1}{2}(\bm{n}\cdot\text{curl}\bm{n})^2-\frac{1}{2}(\text{div}\bm{n})^2\right]^2+\frac{1}{4}k_5|\bm{n}\times\text{curl}\bm{n}|^4
\nonumber\\
&-&k_6\left[(\bm{n}\cdot\text{curl}\bm{n})\text{curl}\bm{n}\cdot(\nabla\bm{n})(\bm{n}\times\text{curl}\bm{n})+
\frac{1}{2}(\bm{n}\cdot\text{curl}\bm{n})^2|\bm{n}\times\text{curl}\bm{n}|^2\right].
\end{eqnarray}
To be more specific, we first reduce the general Euler-Lagrange equations by looking for solutions which are invariant under translations along the $\bm{e}_x$ and $\bm{e}_y$ directions, namely
\begin{equation}
\label{ansatz_new}
\bm{n}= \sin\theta(z)\cos\phi(z)\bm{e}_x+\sin\theta(z)\sin\phi(z)\bm{e}_y+\cos\theta(z)\bm{e}_z.
\end{equation}
Substituting (\ref{ansatz_new}) in \eqref{virga_free_energy2}, we arrive at the reduced form 
\begin{eqnarray}
\label{red_lagrangian}
F_{TB}(\theta,\theta',\phi')&=&\frac{1}{2}\left[k_3\cos^2\theta+(k_1+k_2)\sin^2\theta\right]\theta'^2+\frac{1}{8}(8k_2\sin^4\theta+k_3\sin^22\theta)\phi'^2\nonumber\\
&+&\frac{1}{4}\left(k_5\cos^4\theta+k_4\sin^4\theta\right)\theta'^4\nonumber\\
&+&\frac{1}{8}\left(-4k_6\cos^2\theta\sin^6\theta+4k_4\sin^8\theta+\frac{1}{8}k_5\sin^42\theta\right)\phi'^4\nonumber\\
&+&\frac{1}{2}\left(k_5\cos^4\theta\sin^2\theta-k_6\cos^2\theta\sin^4\theta+k_4\sin^6\theta\right)\theta'^2\phi'^2,
\end{eqnarray}
where $^\prime$ denotes derivative with respect to $z$.

The corresponding reduced Euler-Lagrange equations associated with (\ref{red_lagrangian}) read as
\begin{eqnarray}
\label{phi_eq_2}
&&\Bigg[\frac{1}{2}\phi'^3\left(-4k_6\cos^2\theta\sin^6\theta+4k_4\sin^8\theta+\frac{1}{8}k_5\sin^42\theta\right)
+\frac{1}{4}\phi'\left(8k_2\sin^4\theta+k_3\sin^22\theta\right)\nonumber\\
&&+\phi'\theta'^2\left(k_5\cos^4\theta\sin^2\theta-k_6\cos^2\theta\sin^4\theta+k_4\sin^6\theta\right)\Bigg]'=0,
\end{eqnarray}
and
\begin{eqnarray}
\label{theta_eq_2}
&&\Bigg[\frac{1}{2}\theta'\left[2k_3\cos^2\theta+2(k_1+k_2)\sin^2\theta\right]+\theta'^3\left(k_5\cos^4\theta+k_4\sin^4\theta\right)\nonumber\\
&+&
\theta'\phi'^2\left(k_5\cos^4\theta\sin^2\theta-k_6\cos^2\theta\sin^4\theta+k_4\sin^6\theta\right)\Bigg]'\nonumber\\
&-&\frac{1}{2}\theta'^2\frac{d}{d\theta}\left[k_3\cos^2\theta+(k_1+k_2)\sin^2\theta\right]
-\frac{1}{8}\phi'^2\frac{d}{d\theta}(8k_2\sin^4\theta+k_3\sin^22\theta)-2\theta'^4\frac{d}{d\theta}\left(k_5\cos^4\theta+k_4\sin^4\theta\right)\nonumber\\
&-&\frac{1}{8}\phi'^4\frac{d}{d\theta}\left(-4k_6\cos^2\theta\sin^6\theta+4k_4\sin^8\theta+\frac{1}{8}k_5\sin^42\theta\right)\nonumber\\
&-&\frac{1}{2}\theta'^2\phi'^2\frac{d}{d\theta}\left(k_5\cos^4\theta\sin^2\theta-k_6\cos^2\theta\sin^4\theta+k_4\sin^6\theta\right)=0.
\end{eqnarray}
Since we are interested first in solutions with constant $\theta$, it is straightforward to verify that $\theta=\theta_0$ and $\phi(z)=\beta z$, together with the trivial nematic configuration, solve simultaneously  equations \eqref{phi_eq_2} and \eqref{theta_eq_2} which turn into equations for $\theta_0$ and $\beta$. Solving the latter equations is equivalent 
 to solving the stationary conditions with respect to $\theta_0$ and $\beta$
of the deflated free-energy density 
\begin{eqnarray}
F_{TB}(\bm{n}_h)=f_{TB}(\theta_0,\beta)&=&\frac{1}{8}\left(-4k_6\cos^2\theta_0\sin^6\theta_0+4k_4\sin^8\theta_0+\frac{1}{8}k_5\sin^42\theta_0\right)\beta^4\nonumber\\
&+&\frac{1}{8}\left(8k_2\sin^4\theta_0+k_3\sin^22\theta_0\right)\beta^2.
\end{eqnarray}
Upon setting $t:=\sin^2\theta_0$ we arrive at
\begin{eqnarray}
f_{TB}(t,\beta)=\frac{1}{4}\left[-2k_6(1-t)t+2k_4t^2+k_5(1-t)^2\right]t^2\beta^4+\frac{1}{2}\left[2k_2t+k_3(1-t)\right]t\beta^2.
\end{eqnarray}
The stationary conditions
\begin{equation}
2\sin\theta_0\cos\theta_0\frac{\partial f_{TB}}{\partial t}=0,\qquad\quad \frac{\partial f_{TB}}{\partial\beta}=0.
\end{equation}
correspond to the Euler-Lagrange equations above and they read as
\begin{eqnarray}
\beta^2\Bigg\{k_3+4k_2t-2k_3t+\beta^2t\Bigg[k_5\Bigg(1-3t+2t^2\Bigg)+t\Bigg(4k_4t+k_6\Bigg(-3+4t\Bigg)\Bigg)\Bigg]\Bigg\}=0,
\end{eqnarray}
\begin{eqnarray}
\beta t\Bigg\{k_3+2k_2t-k_3t+\beta^2t\Bigg[k_5(1-t)^2+2t\Bigg(k_6(t-1)+k_4t\Bigg)\Bigg]\Bigg\}=0.
\end{eqnarray}
The only non-trivial solutions are the heliconical configurations given by
\begin{eqnarray}
\label{t}
t=\frac{2k_2k_5+k_3k_6}{(2k_2k_5+k_3k_6)+2(k_3k_4+k_2k_6)},
\end{eqnarray}
\begin{equation}
\label{beta}
\beta=\pm\frac{(2k_2k_5+k_3k_6)+2(k_3k_4+k_2k_6)}{\sqrt{-(2k_2k_5+k_3k_6)(2k_4k_5-k_6^2)}},
\end{equation}
from which we can obtain the conical angle 
\begin{equation}\label{def_theta_0}
\theta_0=\arcsin{\left(\sqrt{\frac{2k_2k_5+k_3k_6}{(2k_2k_5+k_3k_6)+2(k_3k_4+k_2k_6)}}\right)}.
\end{equation}

Now we show that the heliconical solution is a local minimum of the free-energy density by the analysis of the Hessian matrix
\begin{equation}
H=\left(\begin{matrix}
H_{tt} & H_{t\beta}\\
H_{t\beta} & H_{\beta\beta}
\end{matrix}
\right),
\end{equation}
where
\begin{equation}
H_{tt}=\frac{\partial^2f}{\partial t^2}= \frac{N_{tt}}{D_{tt}},\qquad H_{\beta\beta}=\frac{\partial^2f}{\partial \beta^2},
\qquad H_{t\beta}= H_{\beta t}=\frac{\partial^2f}{\partial t\partial \beta},
\end{equation}
with
\begin{eqnarray}
&&N_{tt}=\nonumber\\
&&\Big(4 (k_3 (2 k_4 + k_6) + 
   2 k_2 (k_5 + k_6))^2 (4 k_2^2 k_5 (8 k_4 k_5 + k_5^2 + 2 k_5 k_6 - 3 k_6^2) + 
   k_3^2 (4 k_4^2 k_5 - 4 k_4 k_5 k_6 + k_6^2 (k_5 + 4 k_6))\nonumber\\
&&+ 
   4 k_2 k_3 (-2 k_4 k_5 (k_5 - 3 k_6) + 
      k_6 (k_5^2 + 3 k_5 k_6 - 2 k_6^2)))\Big),
\end{eqnarray}
\begin{eqnarray}
D_{tt} = 8((2 k_2 k_5 + k_3 k_6)^2 (-2 k_4 k_5 +
    k_6^2)^2),
\end{eqnarray}
\begin{equation}
H_{\beta\beta}=-\frac{4 (2 k_2 k_5 + k_3 k_6) (k_3^2 k_4 + 2 k_2^2 k_5 + 
    2 k_2 k_3 k_6)}{(k_3 (2 k_4 + k_6) + 2 k_2 (k_5 + k_6))^2},
\end{equation}
\begin{equation}
H_{t\beta}= \frac{8 k_2^2 k_5 - 2 k_2 k_3 (k_5 - 3 k_6) + 
  k_3^2 (2 k_4 - k_6)}{\sqrt{[-(2 k_2 k_5 + k_3 k_6) (2 k_4 k_5 - k_6^2)]}},
\end{equation}
\begin{equation}
\text{Det}H=-\frac{(2 k_2 k_5 + k_3 k_6) (k_3 (2 k_4 + k_6) + 2 k_2 (k_5 + k_6))^2}{(-2 k_4 k_5 + 
   k_6^2)^2}.
\end{equation}
By considering the known constraints on the elastic constants (see main text), it is trivial to see that $H_{\beta\beta}>0$ and  $\text{Det} H>0$,
and therefore $\bm{n}_h$ turns out to be locally stable.


\subsection{Non-Uniform localised states: Asymptotics}

In this section, we report details about the asymptotics ($r=0$ and $r \rightarrow \infty$) of the Euler-Lagrange equation for localised solutions with $\alpha = 1$.

In order to study the behaviour around $r=0$ we first fix the leading order power at the origin by assuming
that, close to $r=0$, the profile function $f$ takes the form 
\begin{equation}
\label{ex_78}
f(r)=ar^l+\text{O}(r^{l+1}),
\end{equation}
with $l>0$, as a negative value would imply loss of regularity in $f$ at the origin.
In addition, $l$ has to take an odd value due to the symmetry 
\begin{equation}
\label{symmetry}
f\rightarrow -f\quad\text{and}\quad r\rightarrow -r.
\end{equation}

Here we show that $l=1$. To this aim, we expand the l.h.s. term of the Euler-Lagrange equation around $f=0$ which is the value taken by $f$ at $r=0$.
By introducing the above expansions in the functions $G_0$, $G_1$, $G_2$, $G_3$ and $G_4$ and their derivatives and by replacing $f(r)$ with (\ref{ex_78}),
we obtain
\begin{equation}
\label{ode1:exp2}
\sum_{k=0}^{\infty}{\left[A_{1k}r^{(2k+1)l-3}+A_{2k}r^{(2k+1)l-1}+A_{3k}r^{(2k+3)l-3}+A_{4k}r^{(2k+3)l-1}+A_{5k}r^{(2k+5)l-3}+A_{6k}r^{(2k+1)l+1}\right]}=0,
\end{equation}
with $k\in\mathbb{N}$ and where
\begin{equation}
A_{ik}=A_{ik}(a, l, \beta,k_1,k_2,k_3,k_4,k_5,k_6),\quad\quad i=1,2,3,4,5,6,
\end{equation}
are coefficients arising from the expansion.
According to the symmetry property (\ref{symmetry}),
the above expansion consists of even powers of $r$ only.
Of course, all coefficients of all independent powers have to equate to zero.
Moreover, it can be shown that there are no powers $r^{l-3}$, {\emph i. e.} $A_{10}=0$.
Hence it turns out that the lowest order power is $r^{l-1}$ and the corresponding coefficient in the expansion is 
\begin{equation}
\label{coeff_l-1}
128a(l^2-1)(k_1+3k_2),
\end{equation}
which implies that
\begin{equation}
l=\pm 1,
\end{equation}
thus the only positive root is $l=1$. It also turns out that in (\ref{ode1:exp2}) a power of the form $r^{3(l-1)}$ arises,
which implies that the corresponding coefficient must vanish when $l=1$ in order to annihilate all zero power terms in the expansion.
The corresponding coefficient is
\begin{equation}
\label{coeff_3l-3}
32a^3k_4(-1-66l+96l^2-62l^3+33l^4),
\end{equation}
which does vanish as $l=1$.
As it can be argued from the terms in (\ref{ode1:exp2}), there are no other powers of the form $r^{h(l-1)}$, $h\in \mathbb{N}$.
Thus, we conclude that at the origin $f$ has a linear growth. It is worth noticing that the presence of powers $r^{(l-3)}$ would have implied that
$l=3$ as $l-3$ would have been the lowest order in the expansion. This prevents a cubic growth of $f$ at $r=0$.

Let us now focus on $r\rightarrow\infty$ by considering
\begin{equation}
\label{perturb}
f(r) = f_0+\epsilon h(r),
\end{equation}
where $\epsilon \ll 1$.
By plugging the above perturbation (\ref{perturb}) into the Euler-Lagrange equation,
we obtain a huge expansion in powers of $\epsilon$. 
At the linear order in $\epsilon$
we get the following equation
\begin{eqnarray}
2G_2(r,f_0)h^{\prime\prime}+2\partial_rG_2(r,f_0)h^{\prime}+\left[\partial_{f_0}\partial_{r}G_1(r,f_0)-\partial_{f_0}^2G_0(r,f_0)\right]h=0.
\end{eqnarray}
At the lowest order, by neglecting in the functions $G_i$ all those terms going like $\frac{1}{r}$, $\frac{1}{r^2}$, $\frac{1}{r^3}$ and keeping only the linear terms in $r$, we get the following approximate equation for $h$
\begin{equation}
rh^{\prime\prime}+h^{\prime}-\omega^2hr=0,
\end{equation}
where $\omega^2$ is a parameter depending on elastic constants
\begin{equation}
\omega^2 = \frac{\omega^2_n}{\omega^2_d},
\end{equation}
with
\begin{eqnarray}
\omega^2_n &=& 4 (k_2 k_6 + k_3 k_4) (2 k_2 (k_5 + k_6)+ k_3 (2 k_4+ k_6)) \Big(4 k_2^2 k_5 (8 k_4 k_5+(k_5 - k_6) (k_5+3 k_6))\nonumber\\
&+&4 k_2 k_3 \left(k_6 \left(k_5^2+3 k_5 k_6-2 k_6^2\right)-2 k_4 k_5 (k_5-3 k_6)\right) + k_3^2 \left(4 k_4^2 k_5-4 k_4 k_5 k_6 + k_6^2 (k_5+4 k_6)\right)\Big),
\end{eqnarray}
\begin{eqnarray}
\omega^2_d &=& \left(2 k_4 k_5 - k_6^2\right) (2 k_2 k_5 + k_3 k_6) 
\Big(-k_1 \left(2 k_4 k_5 - k_6^2\right) (k_2 k_6 + k_3 k_4)\nonumber\\
&+&k_2^2 \left(8 k_4 k_5 (k_5 + k_6) + k_6 \left(4 k_5^2+8 k_5 k_6+5 k_6^2\right)\right)\nonumber\\
&+&2 k_2 k_3 \left(4 k_4^2 k_5 + k_4 k_6 (8 k_5+7 k_6)+2 k_6^2 (k_5 + k_6)\right) + k_3^2 k_6 (3 k_4 + k_6)^2\Big).
\end{eqnarray}
If $\omega^2>0$, corresponding to the cases of interest (see the elastic constant constraints),
we can rescale $r$ by setting $x:=\omega r$ and ending up with
\begin{equation}
x\frac{d^2h}{dx^2}+\frac{dh}{dx}-xh=0,
\end{equation}
which is a modified Bessel equation with general solution
\begin{equation}
h=c_1I_0(x)+c_2K_0(x),
\end{equation}
where $I_0$ and $K_0$ are modified Bessel functions of first and second kind and order zero, with $c_1$ and $c_2$ two arbitrary constants.
The boundary condition $h\rightarrow0$ as $r\rightarrow \infty$ provides
\begin{equation}
h(r)=c_2K_0(\omega r)\approx c_2\sqrt{\frac{\pi}{2}}\frac{e^{-\omega r}}{\sqrt{\omega r}}+...\;.
\end{equation}
as $I_0$ diverges at infinity.


\subsection{Normal form}
The Euler-Lagrange differential equation for case $\alpha=1$ can be written in normal form as follows
\begin{equation}
\label{N_D}
f^{\prime\prime}=\frac{N}{D},
\end{equation}
where
\begin{eqnarray}
\label{N_normal_f}
N = -\left[2f^{\prime}\partial_rG_2+f^{\prime 2}\partial_fG_2+2f^{\prime 3}(\partial_fG_3+2\partial_rG_4)+3f^{\prime 4}\partial_fG_4-\partial_fG_0+\partial_rG_1\right],
\end{eqnarray}
and
\begin{equation}
\label{D_normal_f}
D=2(G_2+3f^{\prime}G_3+6f^{\prime 2}G_4).
\end{equation}
More precisley, we can rewrite the equation as
\begin{equation}
\label{N_D_exp}
f^{\prime\prime}=-\frac{\left[2f^{\prime}H_{2r}+f^{\prime 2}H_{2f}+2f^{\prime 3}(H_{3f}+2H_{4r})+3f^{\prime 4}H_{4f}-H_{0f}+H_{1r}\right]}{2(H_2+3f^{\prime}H_3+6f^{\prime 2}H_4)},
\end{equation}
where the quantities defining the r.h.s. term are collected below.
\begin{equation}
H_{2r} = h_{r21}+h_{r22}\cos 2f+h_{r23}\cos 4f+h_{r24}\cos 6f,
\end{equation}
with
\begin{eqnarray}
h_{r21}=-r(71k_4+5k_5+29k_6)+4r^3\left[8(k_1+5k_2+k_3)+\beta^2(37k_4+k_5+9k_6)\right],
\end{eqnarray}
\begin{equation}
h_{r22}=\frac{r}{2}(15k_4+15k_5+79k_6)+2r^3\left[16(k_1+k_2-k_3)-\beta^2(97k_4+k_5+17k_6)\right],
\end{equation}
\begin{eqnarray}
h_{r23}=-r(-63k_4+3k_5+11k_6)+4\beta^2r^3(11k_4-k_5-k_6),
\end{eqnarray}
\begin{eqnarray}
h_{r24}=(k_4+k_5+k_6)(2\beta^2r^3+\frac{r}{2}).
\end{eqnarray}
As for $H_{2f}$
\begin{equation}
H_{2f} = -2h_{22}\sin 2f-4h_{23}\sin 4f-6h_{24}\sin 6f,
\end{equation}
where
\begin{eqnarray}
h_{22}=-\frac{r^2}{2}(15k_4+15k_5+79k_6)+2r^4\left[16(k_1+k_2-k_3)-\beta^2(97k_4+k_5+17k_6)\right],
\end{eqnarray}
\begin{equation}
h_{23}=r^2(-63k_4+3k_5+11k_6)+4\beta^2r^4(11k_4-k_5-k_6),
\end{equation}
\begin{eqnarray}
h_{24}=(k_4+k_5+k_6)(2\beta^2r^4-\frac{r^2}{2}).
\end{eqnarray}
As for $H_{3f}$
\begin{eqnarray}
H_{3f}=2h_{31}\cos 2f+4h_{32}\cos 4f,
\end{eqnarray}
where
\begin{eqnarray}
h_{31}=-8r^3(6k_4+k_6),\qquad\qquad h_{32}=-4r^3(4k_4-k_6).
\end{eqnarray}
As for $H_{4r}$
\begin{equation}
H_{4r}=h_{r41}+h_{r42}\cos 2f+h_{r43}\cos 4f,
\end{equation}
where
\begin{eqnarray}
h_{r41}=r^3(65k_4+9k_5-8k_6),
\end{eqnarray}
\begin{eqnarray}
h_{r42} = 4r^3(5k_4-3k_5+2k_6),
\end{eqnarray}
\begin{eqnarray}
h_{r43} = 3r^3(k_4+k_5).
\end{eqnarray}
As for $H_{4f}$
\begin{eqnarray}
H_{4f}=-2h_{42}\sin 2f-4h_{43}\sin 4f,
\end{eqnarray}
where
\begin{eqnarray}
h_{42} = 4r^4(5k_4-3k_5+2k_6),
\end{eqnarray}
\begin{eqnarray}
h_{43} = 3r^4(k_4+k_5).
\end{eqnarray}
As for $H_{0f}$
\begin{equation}
H_{0f} = -2h_{02}\sin 2f-4h_{03}\sin 4f -6 h_{04}\sin 6f-8h_{05}\sin 8f,
\end{equation}
where
\begin{eqnarray}
&&h_{02}=\\ 
&&-\left[\frac{1}{2}(25k_4+21k_5-3k_6)+32r^2(k_1+k_2+k_3)+2\beta^2r^2(21k_4+3k_5-10k_6)+4\beta^2r^4(32k_2+14\beta^2k_4-k_6\beta^2)\right],
\nonumber
\end{eqnarray}
\begin{eqnarray}
&&h_{03} = \\
&&\frac{1}{4}(2k_4+21k_5+6k_6)-2r^2(8k_2-4k_3+3\beta^2k_5+10\beta^2k_6)+2\beta^2r^4(16k_2-8k_3+14\beta^2k_4-\beta^2k_5+2\beta^2k_6),
\nonumber
\end{eqnarray}
\begin{eqnarray}
&&h_{04}=\\
&&-\left[\frac{1}{2}(-k_4+3k_5+3k_6)-2\beta^2r^2(5k_4+3k_5+6k_6)+4\beta^4r^4(2k_4+k_6)\right],\nonumber
\end{eqnarray}
\begin{eqnarray}
h_{05}=(2k_4+k_5+2k_6)(\frac{3}{16}-\frac{3}{2}\beta^2r^2+\frac{\beta^4}{2}r^4).
\end{eqnarray}
As for $H_{1r}$
\begin{equation}
H_{1r} = h_{r11}\sin 2f + h_{r12}\sin 4f + h_{r13}\sin 6f,
\end{equation}
where
\begin{eqnarray}
&&h_{r11} = -2(-35k_4+5k_6),\\
&&h_{r12} = -8(4k_4-k_6),\\
&&h_{r13} = -2(k_4+k_6).
\end{eqnarray}
As for $H_2$
\begin{equation}
H_2 = h_{21} + h_{22}\cos 2f + h_{23}\cos 4f+h_{24}\cos 6f,
\end{equation}
where
\begin{eqnarray}
h_{21} = r^2(71k_4+5k_5+29k_6)+4r^4\left[8(k_1+5k_2+k_3)+\beta^2(37k_4+k_5+9k_6)\right],
\end{eqnarray}
\begin{eqnarray}
h_{22} = -\frac{1}{2}r^2(15k_4+15k_5+79k_6)+2r^4\left[16(k_1+k_2-k_3)-(97k_4+k_5+17k_6)\beta^2\right],
\end{eqnarray}
\begin{eqnarray}
h_{23 } = r^2(-63k_4+3k_5+11k_6)+4\beta^2r^4(11k_4-k_5-k_6),
\end{eqnarray}
\begin{eqnarray}
h_{24} = (k_4+k_5+k_6)(2\beta^2r^4-\frac{1}{2}r^2).
\end{eqnarray}
As for $H_3$
\begin{eqnarray}
H_3 = h_{31} \sin 2f + h_{32}\sin 4f,
\end{eqnarray}
where
\begin{eqnarray}
&&h_{31} = -8r^3(6k_4+k_6),\\
&&h_{32} = -4r^3(4k_4-k_6).
\end{eqnarray}
As for $H_4$
\begin{equation}
H_4 = h_{41}+h_{42}\cos 2f+h_{43}\cos 4f,
\end{equation}
where
\begin{eqnarray}
&&h_{41} = r^4(65k_4+9k_5-8k_6),\\
&&h_{42} = 4r^4(5k_4-3k_5+2k_6),\\
&&h_{43} = 3r^4(k_4+k_5).
\end{eqnarray}
We can now exploit the normal form to partially verify the symmetry structure of $f(r)$.
First of all, let us assume that $f$ admits a general power expansion around $r=0$ starting from the linear term
\begin{equation}
f(r) = \sum_{k=1}^\infty{\frac{f^{(k)}(0)}{k!}}r^k = \xi r+\upsilon r^2+\zeta r^3+\rho r^4+...\;,
\end{equation}
where $\xi=f^{\prime}(0)$, $\upsilon=\frac{1}{2}f^{\prime\prime}(0)$, $\zeta=\frac{1}{3!}f^{\prime\prime\prime}(0)$, etc... .
Correspondingly, $N$ in (\ref{N_normal_f}) and $D$ in (\ref{D_normal_f}) at the lowest order in $r$ have the following forms
\begin{equation}
N = Ar^4+\text{o}(r^4),
\end{equation}
\begin{equation}
D = A'r^4+\text{o}(r^4),
\end{equation}
where
\begin{equation}
A = -128(k_1+3k_2+9k_4\xi^2)\upsilon,
\end{equation}
and
\begin{equation}
A' = 128(k_1+3k_2+9k_4\xi^2).
\end{equation}
Hence, from (\ref{N_D}) or (\ref{N_D_exp})
\begin{equation}
f^{\prime\prime}(0) = \lim_{r\rightarrow 0}{f^{\prime\prime}(r)}=\lim_{r\rightarrow 0}{\frac{N}{D}}=-\upsilon=-\frac{1}{2}f^{\prime\prime}(0),
\end{equation}
which entails that $f^{\prime\prime}(0)=0$. Similarly, it is possible to show that all the odd derivatives at $r=0$ are zero.
Equivalently, one might assume the symmetry structure from scratch, {\emph{i. e.}}
\begin{equation}
f(r) = \sum_{k=0}^\infty{\frac{f^{(2k+1)}(0)}{(2k+1)!}}r^k = \xi r+\zeta r^3+...\;,
\end{equation}
where, as above, $\xi=f^{\prime}(0)$, $\zeta=\frac{1}{3!}f^{\prime\prime\prime}(0)$. By expanding, correspondingly, $N$ and $D$ in (\ref{N_D}) or (\ref{N_D_exp})
we get
\begin{equation}
N = \tilde{A}r^5+\text{O}(r^7),
\end{equation}
and
\begin{equation}
D = \tilde{B} r^4+\text{O}(r^6),
\end{equation}
where
\begin{equation}
\tilde{A} = -\frac{256}{3}\left[-2(k_1-3k_2)\xi^3+\beta^2(-3k_3\xi+6k_6\xi^3)+3(k_1+3k_2+9k_4\xi^2)\zeta\right],
\end{equation}
and
\begin{equation}
\tilde{B} = 128(k_1+3k_2+9k_4\xi^2).
\end{equation}
Thus,
\begin{equation}
f^{\prime\prime}(0)=\lim_{r\rightarrow 0}{f^{\prime\prime}(r)}=\lim_{r\rightarrow 0}{\frac{N}{D}}=\lim_{r\rightarrow 0}\frac{\tilde{A}}{\tilde{B}}r=0.
\end{equation}
It is also possible to find directly the relationship among the first and third derivatives $f^{\prime}(0)$, $f^{\prime\prime\prime}(0)$ at zero by simply differentiating $f^{\prime\prime}$.
This calculation leads to
\begin{equation}
\zeta = \frac{3\beta^2k_3 \xi+2(k_1-3k_2-3\beta^2k_6) \xi^3}{12(k_1+3k_2+9k_4 \xi^2)},
\end{equation}
which is exactly the same as found in a different way in the text.

\end{document}